\documentclass[11pt,a4paper,fleqn]{article}
\usepackage{jcappub}
\usepackage{graphicx}
\usepackage{hyperref}
\usepackage{subcaption}
\usepackage{amsmath}
\usepackage{multirow}
\usepackage{makecell}
\usepackage{dcolumn}
\usepackage{amssymb}
\usepackage{amsfonts}
\usepackage{amsbsy}

\begin{document}

\title{Prospect of Measuring the Cosmic Dipole by Associating Strongly Lensed Gravitational Waves with Galaxy Surveys}

% \date{\today}

\author[a]{Anson Chen,}
\emailAdd{chena@ucas.ac.cn}  
% \altaffiliation{chena@ucas.ac.cn}
\affiliation[a]{International Center for Theoretical Physics Asia-Pacific, University of Chinese Academy of Sciences, 100190 Beijing, China}

\author[a,b]{Jun Zhang}
%\altaffiliation{Kitt Peak National Observatory}
\affiliation[b]{Taiji Laboratory for Gravitational Wave Universe, University of Chinese Academy of Sciences, 100049 Beijing, China}

\abstract{
    The cosmic dipole observed in the cosmic microwave background (CMB) is traditionally interpreted as being caused by the observer's motion relative to the background. However, tensions with dipole measurements from radio galaxy counts motivate the need for independent probes. This work investigates the feasibility of using strongly lensed gravitational wave (GW) events to measure the cosmic dipole. Strongly lensed GWs produce multiple time-delayed images, which can be used to infer the distances to both the lens and the source. These distances, associated with the observed redshifts of the lens and the source from galaxy catalogues, encode information about the background cosmology and cosmic dipole effects. By reconstructing a statistical sample of doubly lensed GW events based on the singular isothermal sphere lens model, the cosmic dipole can be estimated jointly with background cosmological parameters. Using realistic simulations for Einstein Telescope and Cosmic Explorer, we forecast that a dipole magnitude $g$ consistent with both the CMB and number count measurement could be detected with 10 years of observation. Furthermore, constraints on $g$ are greatly improved by combining constraints from doubly lensed events with those from triply or quadruply lensed events. In the most optimistic scenario, where we measure the number count dipole magnitude with 10 years of observation, we obtain $g = (2.45^{+1.53}_{-1.28}) \times 10^{-3}$ from the combined constraint, provided that systematic uncertainties are mitigated. Although challenging, strongly lensed GWs offer a novel approach to measuring the cosmic dipole, providing an independent consistency test with different systematics from electromagnetic probes.
}

\maketitle

\section{Introduction} 

The cosmic dipole tension is one of the challenges to the Lambda Cold Dark Matter ($\Lambda$CDM) standard cosmology model. Both kinetic motion of the solar system with respect to the background and the intrinsic anisotropy of the universe could cause the observed cosmic dipole due to Doppler effects. An accurate measurement of the cosmic dipole is essential to the test of the cosmology principle of homogeneity and isotropy, and the tests of dark energy theories arising from back reaction of cosmological perturbations \cite{Kolb:2005da,Bolejko:2016qku}. However, two current major methods to estimate the cosmic dipole yield results in tension with each other. 

On one hand, the Cosmic Microwave Background (CMB) temperature map observed by the Planck satellite indicates that the cosmic dipole has an amplitude of $|\vec{v}_o/c|=(1.23\pm0.00036)\times10^{-3}$, corresponding to an observer's velocity of $v_o\sim370~{\rm km~s}^{-1}$ with respect to the CMB rest frame ($c$ is the speed of light), towards a direction of $(264.02^\circ\pm0.01^\circ, 48.253^\circ\pm0.005^\circ)$ \cite{Planck:2018nkj}. On the other hand, one can measure the cosmic dipole by number count of the observed galaxies, because of the observed magnitude changes from the Doppler effect, as well as the angular aberration effect.
% because the observed magnitudes of galaxies are affected by the dipolar Doppler effects, so that the number of galaxies above the observed magnitude threshold would vary in different observed directions. Moreover, the observer kinetic motion also causes angular aberration that changes observed source numbers in different directions. 
The amplitude of the number count dipole is proposed by Ellis and Baldwin \cite{Ellis:1984uka} to be
\begin{equation}
    {\cal D} = [2+x(1+\alpha)]|\vec{v}_o/c|,
\end{equation}
where $x$ is called the magnification bias parameter defined by the slope of the integral source count at the flux limit, which is usually close to 1, and $\alpha$ is the source population spectral index. For radio sources and quasars, the luminosity spectrum obeys $L\propto\nu^{-\alpha}$, with $\nu$ being the emission frequency of the source. However, multiple observations of the number count dipole from various sources yield $|\vec{v}_o|$ about twice of the value measured from CMB, ranging from $600$ to $1000~{\rm km~s}^{-1}$, towards the same direction as the CMB dipole. For instance, Secrest et al. reports a number count dipole of $1.5\times10^{-2}$ with quasars in the CatWISE2020 catalogue, corresponding to a $4.9\sigma$ tension with the CMB dipole \cite{Secrest:2020has}. It is later increased to over $5\sigma$ when including more quasar samples \cite{Secrest:2022uvx,Dam:2022wwh,Land-Strykowski:2025gkz}. In addition, the number count dipole inferred from supernova (SNIa) surveys gives $(5\pm1.6)\times10^{-3}$, which has a $3.3\sigma$ tension between the CMB result \cite{Singal:2021crs}. Furthermore, radio source catalogues like the TIFR GMRT Sky Survey (TGSS), the NRAO VLA Sky Survey (NVSS), and the Westerbork Northern Sky Survey (WENSS), yield a number count dipole from $0.010$ to $0.070$ with an uncertainty in the order of $10^{-3}$ \cite{Blake:2002gx,2011ApJ...742L..23S,Gibelyou:2012ri,Tiwari:2013vff,Fernandez-Cobos:2013fda,Tiwari:2013ima,Bengaly:2017slg,Siewert:2020krp,Oayda:2024hnu,Land-Strykowski:2025gkz}, significantly larger than the dipole from CMB. In spite of some explanations to alleviate the cosmic dipole tension, for example the local structure impact on the local dipole \cite{Tiwari:2015tba,Colin:2017juj,Aluri:2022hzs,Mittal:2023xub}, or the existence of cosmic void in the local universe \cite{Rubart:2014lia}, the tension is not fully reconciled. In addition, the possible impact from time evolution of the radio source population on the Ellis-Baldwin equation has been debated, yet the Ellis-Baldwin equation remains valid \cite{Dalang:2021ruy,vonHausegger:2024jan,vonHausegger:2024fcu}. However, a novel tomographic approach finds a cosmic dipole aligned with the CMB measurement from eBOSS data \cite{daSilveiraFerreira:2024ddn}. Moreover, there are also efforts to measure the intrinsic dipole directly \cite{Ferreira:2020aqa,Ferreira:2021omv}. Nevertheless, more observations on different aspects are required to obtain a better understanding on the cosmic dipole tension.

Meanwhile, the rapid development of gravitational wave (GW) detection in recent years has opened a new window to study cosmology. 
% The first multi-messenger event, GW170817, has provided a new independent measurement of the Hubble constant $H_0$ \cite{LIGOScientific:2017vwq,LIGOScientific:2017zic,LIGOScientific:2017adf}. 
For compact binary coalescence (CBC) events without electromagnetic (EM) counterparts, $H_0$ can be measured using the standard siren approach, which breaks the mass-redshift degeneracy of GW events with features in source mass distribution, and associates GW events with redshifts from galaxy catalogues \cite{Schutz1986,Gray:2019ksv,Gray:2021sew,LIGOScientific:2021aug,Mastrogiovanni:2021wsd,Mastrogiovanni:2023emh,Mastrogiovanni:2023zbw,Gray:2023wgj,Borghi:2023opd,Jin:2023tou,Jin:2025dvf}. The latest GWTC-4 data yields $H_0 = 76.6^{+13.0}_{-9.5} {\rm km~s^{-1}~Mpc^{-1}}$ by combining bright siren constraint from GW170817 and the standard siren measurement with 141 events \cite{LIGOScientific:2025jau}. It is predicted that the precision of standard siren measurement of $H_0$ will reduce to $<1\%$ with next-generation ground-based detectors \cite{Chen:2024gdn}, such as Einstein Telescope (ET) \cite{Branchesi:2023mws} and Cosmic Explorer (CE) \cite{Evans:2021gyd}. In addition, the standard siren method can also constrain modified gravity theories of dark energy via GW propagation effects \cite{Mukherjee:2020mha,Finke:2021aom,Mancarella:2021ecn,Ezquiaga:2021ayr,Leyde:2022orh,Chen:2023wpj,Zhu:2023rrx,Zhang:2024rel,Tagliazucchi:2025ofb,LIGOScientific:2025jau}. 

In addition, GW events are also powerful tools in measuring the cosmic dipole in various frequency bands. First attempts have been made to probe the cosmic anisotropy with LIGO-Virgo-KAGRA (LVK) data via number count or source mass distribution \cite{Stiskalek:2020wbj,Essick:2022slj,Kashyap:2022ibx}, though the current data shows no evidence of anisotropy. However, with $\sim10^6$ GW events detected by ET and CE per year in the future, the cosmic dipole can be measured precisely by number count of CBCs \cite{Mastrogiovanni:2022nya,Grimm:2023tfl}. Apart from the number count method, the cosmic dipole can also be probed accurately by its modification effects on the luminosity distances and redshifts of the events in the era of ET and CE, using $10^3$ bright sirens \cite{Cai:2019cfw,Cousins:2024bhk}, or about 35 golden dark sirens \cite{Chen:2025qsl}. Moreover, massive binary black hole (BBH) mergers as bright sirens will also provide measurements on the cosmic dipole with space-based GW detectors in the millihertz band \cite{Cai:2017aea}. Furthermore, by probing anisotropy in the stochastic GW background, the pulsar timing array can provide constraint on the cosmic dipole as well \cite{Cusin:2022cbb,Tasinato:2023zcg,Cruz:2024svc,Cruz:2024diu}. What's more, the cosmic dipole could potentially be probed by modification to the GW waveform \cite{Cusin:2024git}.

On the other hand, gravitational lensing of GWs has great potential in studying cosmology as well. The strongly lensed GW events will be able to constrain cosmological parameters precisely when associated with EM counterparts \cite{Liao:2017ioi}, or galaxy catalogues \cite{Hannuksela:2020xor,Poon:2024zxn,Chen:2025xeg,Chen:2026htz}. These events could also constrain modified gravity theories via GW propagation effects \cite{Chen:2026htz,Finke:2021znb,Narola:2023viz,Colangeli:2025hrs}. In addition, the population and time-delay distribution of strongly lensed GWs could also provide tight constraints on cosmology and gravity \cite{Jana:2022shb,Jana:2024uta,Maity:2025apt,Ying:2025eem}. Moreover, cosmological parameters could be measured by micro-lensing of GWs \cite{Chen:2024xal}, and cross-correlation between weak-lensing of GWs and galaxy catalogues as well \cite{Mukherjee:2019wcg,Balaudo:2022znx}. With ET and CE deployed, and deeper galaxy surveys by the Dark Energy Spectroscopic Instrument (DESI) \cite{DESI:2025fxa}, the Euclid Mission \citep{Euclid:2024yrr}, and Legacy Survey of Space and Time (LSST) with the Vera C. Rubin Observatory \citep{2009arXiv0912.0201L} in the near future, the association between strongly leansed GWs and galaxy catalogues will be powerful in testing cosmology models. 

In this work, we forecast the prospect to constrain the cosmic dipole by strongly lensed GW events detected by ET and CE associated with galaxy surveys like LSST. By reconstructing the lens model of strongly lensed GW events, we can pin down their host galaxies in the galaxy catalogue. Then we estimate the magnitude and the orientation of the cosmic dipole through the modification in GW luminosity distance and redshift of host galaxies, jointly with $H_0$ and matter energy density today $\Omega_{m,0}$ in the $\Lambda$CDM model. The cosmic dipole will also modify the angular diameter distance and the Einstein radius in the lens system, which are also considered in our lens reconstruction. We also separate doubly lensed events that take up $70\%$ of the total strongly lensed events \cite{Li:2018prc}, and the rest of lensed events with more than two images. While the lens model of doubly lensed events can be approximated by the Singular Isothermal Sphere (SIS) model, the lens model of triply or quadruply lensed events can be more precisely reconstructed using tools like \texttt{lenstronomy} \cite{Birrer:2015rpa}.

The contents of this work are given as the following. First, the strong lensing of GWs is reviewed in Section \ref{sec:GW_lensing}, and then the cosmic dipole effects in strong lensing of GWs are introduced in Section \ref{sec:cosmic_dipole}. Later, we describe our method to probe the cosmic dipole with GW lensing in Section \ref{sec:method}, and our data simulation in Section \ref{sec:simulation}. The results of our forecasts with our method are then given in Section \ref{sec:results}, followed by the conclusions in \ref{sec:conclusions} at the end.

\section{Gravitational wave strong lensing}
\label{sec:GW_lensing}

\subsection{Geometric Optics Approximation}

\begin{figure}[t]
    \centering
    \includegraphics[width=0.7\linewidth]{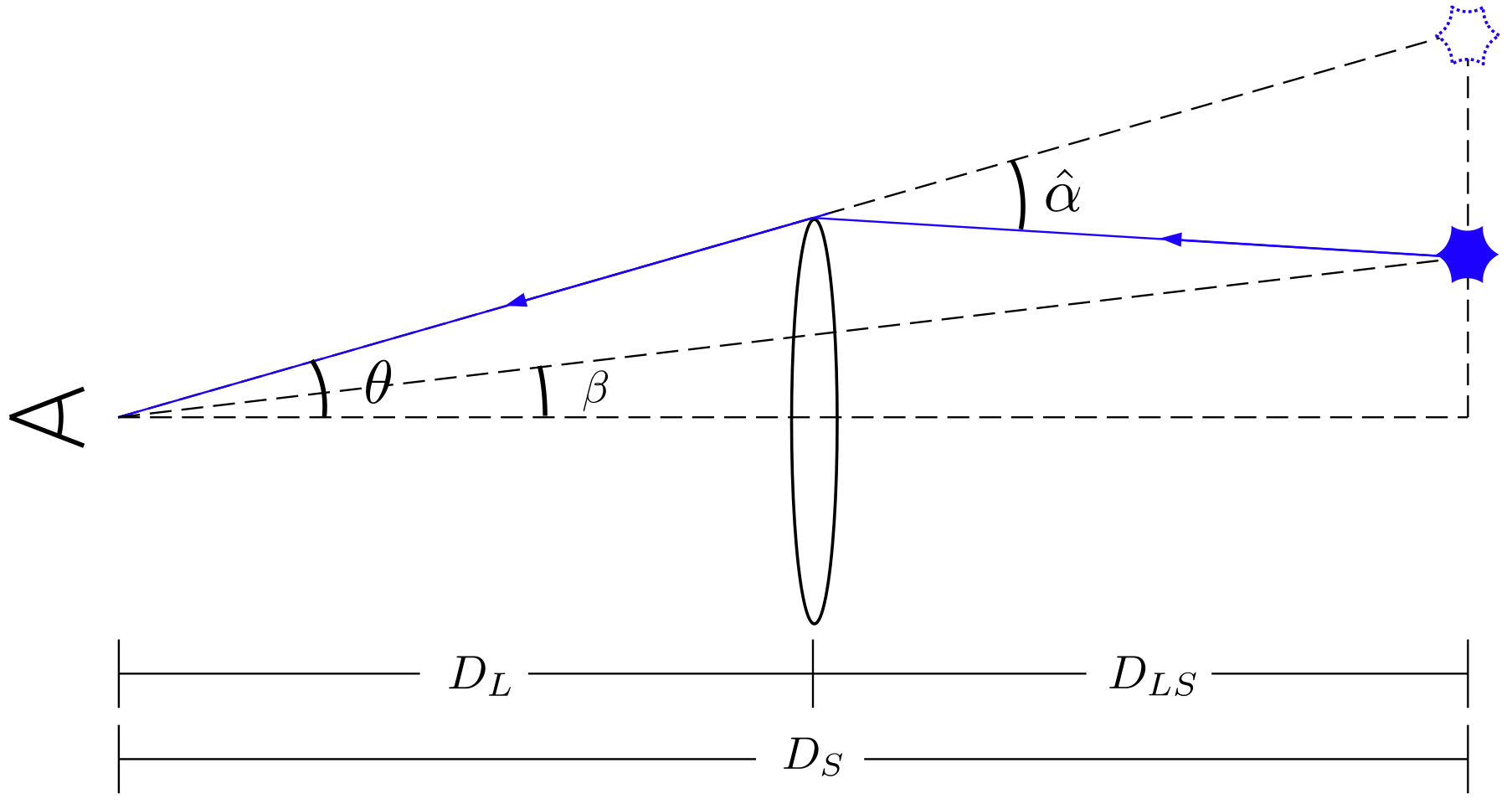}
    \caption{Schematic figure for strong-lensing with the thin-lens approximation. }
    \label{fig:lensing_sketch}
\end{figure}

In this work, we study GWs from CBCs lensed by galaxies on their line-of-sights, in a frequency range of $10-10^3$ Hz for ground-based detectors. In this condition, the wavelength of GWs is much smaller than the size of the lens galaxy, so that we can use the geometric optics approximation for strongly lensed GWs. In the frequency domain, the lensed GW waveform $\tilde{h}_{\rm L}(f)$ is given by
\begin{equation}
    \tilde{h}_{\rm L}(f)=F(\omega)\tilde{h}(f),
\end{equation}
where $F(\omega)$ is the amplification factor, and $\omega=2\pi f$. In the geometric optics regime, $F(\omega)$ is given by \cite{Takahashi_2003,Ezquiaga:2020gdt}
\begin{equation}
    F(\omega) \simeq \sum_j \sqrt{|\mu_j|} \exp\left(i\omega t_{d,j}-i~{\rm sign}(\omega)n_j\frac{\pi}{2} \right),
\end{equation}
where $\mu_j$ is the magnification factor for the $j$-th image of lensed GW signals, $t_{d,j}$ is the time delay of the arrival of the $j$-th image due to lensing, and $n_j=0,1,2$ corresponds to the Morse phase for the type I, II, III image respectively. The time delay of each image of a strongly lensed signal is given by
\begin{equation}
    t_{d,j}(\vec{\theta}_j,\vec{\beta}) = \frac{1+z_L}{c}\frac{D_{\rm L}D_{\rm S}}{D_{\rm LS}}\left[\frac{1}{2}(\vec{\theta}_j-\vec{\beta})^2-\psi(\vec{\theta}_j) \right].
    \label{eq:t_delay}
\end{equation}
Equation (\ref{eq:t_delay}) is constructed in two parts. The first part is caused by the changes in the geometric paths of lensed signal images, where $\vec{\theta}_j$ denotes the angle between the images and the lens on the line-of-sight, and $\vec{\beta}$ denotes the true source angular position (see Fig. \ref{fig:lensing_sketch}). The second part is called the Shapiro time delay, which is caused by the effective gravitational potential of the lens $\psi(\vec{\theta}_j)$. In the pre-factor, $z_L$ denotes the lens redshift, and $D_{\rm L}$, $D_{\rm S}$, and $D_{\rm LS}$ are the angular diameter distances between the lens and the observer, between the source and the observer, and between the lens and the source, respectively.

Equation (\ref{eq:t_delay}) can be further rewritten in dimensionless unit. By scaling the angular position of the image and the source by the Einstein radius $\theta_E$, we have two dimensionless positions, $\vec x=\vec\theta/\theta_E$ and $\vec y=\vec\beta/\theta_E$. The time delay of lensed signals is then given by
\begin{equation}
    t_{d}(\vec{x},\vec{y}) = \frac{1+z_L}{c}\frac{D_{\rm L}D_{\rm S}}{D_{\rm LS}}\theta_E^2 T(\vec x,\vec y),
    \label{eq:t_delay_dimensionless}
\end{equation}
where $T(\vec x,\vec y)$ is the dimensionless time delay defined by 
\begin{equation}
    T(\vec x,\vec y)=\frac{1}{2}(\vec x-\vec y)^2-\Psi(\vec x).
\end{equation}
By choosing a specific spherical symmetric thin-lens model, one can input a specific form of the effective potential $\Psi(\vec x)$, and thus obtain an analytical form of $T(\vec x,\vec y)$.

\subsection{Doubly lensed GWs with the SIS lens model}

Following previous works \cite{Chen:2026htz,Sereno:2011ty}, the lens model of the doubly lensed GW events can be approximated by the SIS model without bringing significant impacts on lens statistic \cite{Maoz:1993ix,Kochanek:1995ap,Mitchell:2004gw}. Although the ellipticity of lens galaxies could lead to biases in cosmological implications using the SIS model, here we want to investigate the detectability of the cosmic dipole by strongly lensed GWs in an ideal condition where lens galaxies have negligible ellipticity.
The SIS model approximates the mass components of the galaxy as ideal gas particles in an adiabatic system \cite{1987gady.book.....B}, so that the mass distribution density profile of the SIS model is given by
\begin{equation}
    \rho(r) = \frac{\sigma_v^2}{2\pi G}\frac{1}{r^2},
\end{equation}
where $\sigma_v$ is the velocity dispersion of stars in the galaxy. One can further infer the Einstein radius of the SIS model to be \cite{Takahashi_2003}
\begin{equation}
    \theta_E = 4\pi\frac{\sigma_v^2}{c^2}\frac{D_{\rm LS}}{D_{\rm S}}.
\end{equation}
On the other hand, the lens mass for the SIS model can be given by \cite{Takahashi_2003}
\begin{equation}
    M_{L} = 4\pi\frac{\sigma_v^4}{Gc^2}\frac{D_{\rm L}D_{\rm LS}}{D_{\rm S}},
\end{equation}
Therefore we have a connection between the Einstein radius and the mass for the SIS model, 
\begin{equation}
\theta_E^2 = 4\pi \frac{GM_L}{c^2}\frac{D_{\rm LS}}{D_{\rm L}D_{\rm S}}.
\label{eq:theta_E_sis}
\end{equation}
More importantly, the SIS model results in a special lens equation. In general, the lens equation describes the relation among the image position, the source position, and the reduced deflection angle by $\vec\beta=\vec\theta-\vec\alpha(\vec\theta)$, where $\alpha$ is defined by $\alpha=\hat{\alpha}D_{LS}/D_S$ ($\hat{\alpha}$ is the deflection angle shown in Fig. \ref{fig:lensing_sketch}). In the SIS model, $\alpha(\theta)=\theta_E$, so that the lens equation yields
\begin{equation}
    \theta_{\pm} = \beta \pm \theta_E,
    \label{eq:sis_lens_eq}
\end{equation}
where $\theta_\pm$ represents the two image position angles in the case of $\beta<\theta_E$. Otherwise, if $\beta\geq\theta_E$, only one image may exist. Hence, for doubly lensed galaxy images, one can infer the Einstein radius for the SIS lens model by $\theta_E=(\theta_+ - \theta_-)/2$ with the image angular positions from observations.

In addition, the dimensionless time delay and the magnification factor of the SIS model are given by \cite{Takahashi:2016jom}
\begin{align}
    T_\pm & = \mp y - \frac{1}{2},\\
    \label{eq:sis_time_delay}
    \mu_\pm & = 1 \pm \frac{1}{y}.
\end{align}
By jointly analyze the two images of strongly lensed GW signals with the SIS model, we can jointly estimate the true luminosity distance of GW events and the impact parameter $y$ \cite{Chen:2026htz}. Then we can infer the analytical time delay with $y$ from GW parameter estimation (PE) and $\theta_E$ from lensed galaxy images, and constrain cosmological parameters by comparing the analytical time delay with the observed time delay.
Compared to the time delay measurement for strongly lensed EM sources \cite{H0LiCOW:2019pvv}, the time delay measurement for strongly lensed GW signals would be much more precise due to small uncertainty in the arrival time of GW signals. The main uncertainty of the analytical time delay will come from the estimation of $y$.

\subsection{Lensed GWs with more than two images}

Apart from doubly lensed events, strongly lensed GW events with 3 to 5 images are also expected to be observed in the near future \cite{Hannuksela:2020xor,Poon:2024zxn,Chen:2025xeg,Uronen:2024bth}. More complex lens models are required to perform lens reconstruction for such events, such as the Singular Isothermal Ellipsoid model \cite{Kormann:1994}, using specific software packages like \texttt{lenstronomy} \cite{Birrer:2015rpa}. The effective potential in strong-lensing time delay can be precisely constructed for these events, so that the time delay between any images is given by \cite{Liao:2017ioi,Chen:2025xeg}
\begin{equation}
    \Delta t_{ij} = \frac{1+z_L}{c}D_{\Delta t}\Delta \phi_{ij},
    \label{eq:t_delay_quad}
\end{equation}
where $D_{\Delta t} = D_{\rm L}D_{\rm S}/D_{\rm LS}$, and $\Delta \phi_{ij}$ is the difference between the effective potential of images. With observed $\Delta t_{ij}$ and reconstructed $\Delta \phi_{ij}$, one can then constrain cosmological parameters by constructing $D_{\Delta t}$ with $z_L$, $z_S$ from galaxy catalogues.

\section{Cosmic dipole modification}
\label{sec:cosmic_dipole}

The cosmic dipole induced by the kinetic motion of the observer with velocity $\vec{v}_o$ in the CMB rest frame can be denoted by $g\cdot\hat{n}=-\vec{v}_o/c$ , where $g$ and $\hat{n}$ represent the magnitude and the direction of the dipole. The dipole direction in the sky coordinate can be further written by 
\begin{equation}
    \hat{n} = (\cos\phi^{\rm dip}\sin\theta^{\rm dip}, \sin\phi^{\rm dip}\sin\theta^{\rm dip}, \cos\theta^{\rm dip}),
\end{equation}
where $\phi^{\rm dip}$ and $\theta^{\rm dip}$ are the right ascension and the declination of the dipole direction. Given a line-of-sight angular position of a source on the sky, denoted by $\hat{z}(\phi,\theta)$, the corresponding magnitude of the cosmic dipole effect is given by $g(\hat{n}\cdot\hat{z})$. As a consequence, the observed luminosity distance of the source is modified by the cosmic dipole effects as \cite{Bonvin:2005ps}
\begin{equation}
    d_L' = d_L(z)[1+g(\hat{n}\cdot\hat{z})],
    \label{eq:DL_dip}
\end{equation}
and the observed redshift of the source is also modified as
\begin{equation}
    1+z' = (1+z)[1+g(\hat{n}\cdot\hat{z})],
    \label{eq:z_dip}
\end{equation}
where $d_L$ and $z$ are the luminosity distance and the redshift in the CMB rest frame. In the late universe within the $\Lambda$CDM framework, we have 
\begin{equation}
	d_L(z) = (1+z)r(z),
\end{equation}
where $r$ is the comoving distance given by
\begin{equation}
    r(z)=\int_0^{z} \frac{c}{H(\tilde{z})}{\rm d}{\tilde z}.
\end{equation}
The Hubble parameter is given by 
\begin{equation}
    H(z)=H_0\sqrt{\Omega_{m,0}(1+z)^3+\Omega_{\Lambda,0}}.
\end{equation}
Here $\Omega_{m,0}$ and $\Omega_{\Lambda,0}$ are energy density of matter and dark energy today. Meanwhile, the observed component masses of CBC GW sources $m^{\rm obs}$ are redshifted from the source masses $m^{\rm s}$, and are thus modified by the cosmic dipole effects as
\begin{equation}
    m' = m^{\rm s}(1+z') = m^{\rm s}(1+z)[1+g(\hat{n}\cdot\hat{z})].
    \label{eq:m_dip}
\end{equation}
In terms of observation, the relation between the observed luminosity distance and the observed redshift is given by \cite{Dalang:2021ruy}
\begin{equation}
    d_L' = d_L(z')\left[1-\frac{c(1+z')g(\hat{n}\cdot\hat{z})}{r(z')H(z')} \right].
    \label{eq:dL_obs_z}
\end{equation}
This is the equation we should use for the standard siren analysis in the presence of the cosmic dipole.

On the other hand, to correctly reconstruct the lens model in strong lensing, we need to construct the true angular diameter distance in the CMB rest frame with the observed redshift of galaxies. As derived in \cite{Dalang:2023usx}, the rest-frame angular diameter distance of the source is given by
\begin{align}
    D_S & = \frac{c}{1+z_S(z_S')} \int_0^{z_S(z_S')}\frac{d\tilde{z}}{H(\tilde{z})} \nonumber\\
    & = \frac{c}{1+z_S(z_S')} \int_0^{z_S'-(1+z_S')g(\hat{n}\cdot\hat{z})}\frac{d\tilde{z}}{H(\tilde{z})} \nonumber\\
    & \simeq D_A(0,z_S')[1+g(\hat{n}\cdot\hat{z})] - \frac{c}{H(z_S')}g(\hat{n}\cdot\hat{z}).
    \label{eq:DS_z_obs}
\end{align}
Here $D_A$ is the rest-frame formalism for angular diameter distance between two redshifts, defined by 
\begin{equation}
    D_A(z_1,z_2)=\frac{c}{(1+z_2)}\int_{z_1}^{z_2} \frac{c}{H(\tilde{z})}{\rm d}\tilde{z}.  
\end{equation}
Similarly, we have
\begin{equation}
    D_L \simeq D_A(0,z_L')[1+g(\hat{n}\cdot\hat{z})] - \frac{c}{H(z_L')}g(\hat{n}\cdot\hat{z}),
    \label{eq:DL_z_obs}
\end{equation}
\begin{equation}
    D_{LS} \simeq D_A(z_L',z_S')[1+g(\hat{n}\cdot\hat{z})] - \frac{c}{H(z_S')}g(\hat{n}\cdot\hat{z}) + \frac{1+z_L'}{1+z_S'}\frac{c}{H(z_L')}g(\hat{n}\cdot\hat{z}).
    \label{eq:DLS_z_obs}
\end{equation}
Equation (\ref{eq:DS_z_obs})-(\ref{eq:DLS_z_obs}) will be used to compute angular diameter distances in lens model reconstruction with the observed redshift of the lens and the source galaxy.

On the other hand, the time dilation effects due to the observer's velocity on the strong lensing time delay is in the order of $v_o^2/c^2$, so that it can be negligible compared to other effects \cite{Dalang:2023usx}. In addition, due to the angular aberration effect of the cosmic dipole, the EM image position angles are also modified. However, the changes in the angle depends on both the line-of-sight direction of the lens and the rotation of the image around the line-of-sight axis \cite{Dalang:2023usx}. Accounting for such an effect requires more complex simulation of the image positions relative to the lens on the sky, which is beyond the scope of this work. In our simulation, we only create random values of the impact parameter $y$ for lensed events without specifying the rotation angle relative to the line-of-sight axis of the lens. Therefore we didn't include the angular aberration effect in our measurement of the cosmic dipole, but it will need to be studied specifically in future's work.

\section{Method}
\label{sec:method}

Following our previous work \cite{Chen:2026htz}, we assume that in the XG-detector era, the precise localization of strongly lensed GWs by three XG detectors will allow us to identify their host galaxies. Especially, for triply lensed or quadruply lensed events, the localization area can be narrow down to $\Delta\Omega\sim0.01~{\rm deg}^2$, and only 1 strong lensing system is expected in the area. For doubly lensed events, the localization would be less accurate, but the true host galaxies can be found by comparing the Bayes factor of the likelihood between different strong lensing systems in the galaxy catalogue within the localization area \cite{Hannuksela:2020xor,Wempe:2022zlk}. In our simulation, we consider the optimistic scenario in which true strongly lensed systems are identified for GW lensed events. In this case, the systems must lie within the footprints of galaxy surveys, and the angular positions of the lensed images must be resolvable by electromagnetic telescopes (see Section \ref{sec:lens_select} for details on lensed event selection). Otherwise, lensed GW events whose host galaxy EM images fall outside existing galaxy catalogues cannot be analyzed using our method.

Since the cosmic dipole effect changes the observed luminosity distances $d_L^{\rm obs}$ of GWs and observed redshift $z^{\rm obs} $ of galaxies, the dipole magnitude can be measured with the standard siren method, along with other cosmological parameters such as $H_0$ and $\Omega_{m,0}$ in the $\Lambda$CDM background. However, we need to remove the magnification by strong lensing of GWs first to obtain true $d_L^{\rm obs}$. Thus, a re-analysis of the lensed GWs is required once the lensed signal pairs are confirmed. Through a joint re-analysis of both GW images of a doubly lensed event, we can obtain posterior samples for true $d_L^{\rm obs}$ and the impact parameter $y$ with the SIS model \cite{Chen:2026htz}. For further details on the re-analysis of waveform parameters, see Section \ref{sec:waveform_PE}. While the doubly lensed events take up $\sim 70\%$ of the total lensed events, the rest of the events contain more than two images, allowing more precise lens reconstruction and thus de-magnification of the signals. Here we assume that the uncertainty of $d_L^{\rm obs}$ for triply or quadruply lensed events is at the same level as doubly lensed events after joint analysis of the images.

On the other hand, the cosmic dipole effect also needs to be considered in angular diameter distances for time delay cosmography with strong lensing, from which we can also constrain the cosmic dipole. For the doubly lensed events, using the observed time delay between GW signal images, $\Delta t = t_{d,2} - t_{d,1}$, along with the redshifts of the lens galaxy and the source galaxy, the Einstein radius $\theta_E$, and the impact parameter $y$, we can estimate cosmological parameters via equations (\ref{eq:t_delay_dimensionless}) and (\ref{eq:sis_time_delay}) under the SIS model. For a resolved image of a strong-lensing system in galaxy surveys, the angular separation between the two images of the source can be measured. From the lens equation (\ref{eq:sis_lens_eq}), the Einstein radius $\theta_E$ of the SIS lensing system can then be obtained as $\theta_E = (\theta_+ - \theta_-)/2 = \Delta\theta/2$. We therefore construct the log-likelihood for the cosmological parameters $\lambda$ using $\chi^2$ statistics over $N$ lensed events combining the standard siren and the time delay measurement:
\begin{equation}
\log{\cal L}(\lambda) = -\frac{1}{2}\sum_{i=1}^N \chi^2(\lambda,\Delta t_i^{\rm obs},\Delta\theta_i,z_{L,i}',z_{S,i}',y_i),
\label{eq:likelihood}
\end{equation}
where
\begin{equation}
\chi^2(\lambda,\Delta t_i^{\rm obs},\Delta\theta_i,z_{L,i}',z_{S,i}',y_i) = \left[\frac{d_{L,i}^{\rm obs}-d_L'(\lambda,z_{S,i}')}{\sigma(d_{L,i}^{\rm obs})} \right]^2 + \left[\frac{\Delta t_i^{\rm obs}-\Delta t(\lambda,\Delta\theta_i,z_{L,i}',z_{S,i}',y_i)}{\sigma(\Delta t)} \right]^2.
\label{eq:chi_square}
\end{equation}
Note that the angular diameter distances $D_S$, $D_L$ and $D_{LS}$ used in analytical time delay construction are computed with equations (\ref{eq:DS_z_obs}), (\ref{eq:DL_z_obs}) and (\ref{eq:DLS_z_obs}) respectively using observed redshift $z_S'$ and $z_L'$. Because of the negligible uncertainty in GW arrival time, which is on the order of milliseconds, compared to strong-lensing time delays of days, the uncertainty $\sigma(\Delta t)$ comes from analytical time delay reconstruction. In particular, it is dominated by $\sigma_y$ from the GW re-analysis, which typically yields a relative error of $5\%–10\%$. The uncertainties in galaxy redshift, $z_L'$ and $z_S'$, is in the order of $1\%-3\%$ in photometric surveys like LSST \cite{2009arXiv0912.0201L}, and it can be further lower to order of $0.1\%$ with follow-up spectroscopic measurements, e.g. Euclid \cite{Euclid:2024yrr} and MegaMapper \cite{Schlegel:2022vrv}, so we assume that the uncertainty in redshift is negligible in ideal measurements. Therefore, we can simply use $\sigma(\Delta t)/\Delta t \simeq \sigma_y/y$. Furthermore, we are assuming a perfectly smooth lens model, so we don't consider the time delay changes from sub-structures of galaxies.
Finally, the posterior for $\lambda$ can be obtained by Bayesian inference with the log likelihood and a flat prior on $\lambda$ using Markov Chain Monte Carlo (MCMC) sampling.

However, unlike doubly lensed events, the effective potential $\Delta \phi$ can be obtained from lens reconstruction with EM images more accurately for triply or quadruply lensed events. Therefore, from equation (\ref{eq:t_delay_quad}), the time delay part in the log likelihood in equation (\ref{eq:chi_square}) simply becomes
\begin{equation}
    \chi^2_{\Delta t}=\left[\frac{\Delta t_i^{\rm obs}/\Delta \phi_i - (1+z_{L,i}')D_{\Delta t}(\lambda,z_{L,i}',z_{S,i}')/c}{\sigma (\Delta \phi_i,z_{L,i}',z_{S,i}')} \right]^2,
    \label{eq:chi_square_quad}
\end{equation}
where the total uncertainty $\sigma$ consists of error from $\Delta \phi$ and galaxy redshifts, which is computed by
\begin{align}
    \sigma^2(\Delta \phi_i,z_{L,i}',z_{S,i}') &= \bigg(\frac{\Delta t_i^{\rm obs}}{\Delta \phi_i}\bigg)^2 \bigg(\frac{\sigma_{\Delta \phi_i}}{\Delta \phi_i}\bigg)^2 \nonumber\\ 
    &+ \bigg(\frac{1+z_{L,i}'}{c}\frac{\partial D_{\Delta t}}{\partial z_{L,i}'}+\frac{D_{\Delta t}}{c} \bigg)^2 \sigma_{z_L'}^2 \nonumber\\ 
    &+ \bigg(\frac{1+z_{L,i}'}{c}\frac{\partial D_{\Delta t}}{\partial z_{S,i}'} \bigg)^2 \sigma_{z_S'}^2.
\end{align}
The relative error of the effective potential $\sigma_{\Delta \phi}/\Delta \phi$ can be reasonably taken as $2\%$, accounting for uncertainties from lens reconstruction and line-of-sight mass distribution \cite{Liao:2017ioi,H0LiCOW:2016xpx,Chen:2025xeg}. We also assume galaxy redshift error to be $\sigma_{z'}=0.001(1+z')$ for ideal spectroscopic measurement.

\section{Simulation}
\label{sec:simulation}

In this work, we forecast the constraint on the cosmic dipole from mock strongly lensed GW data based on realistic models of GW source population, merger rate, and lensing rate. We select the events that can be associated with EM strong-lensing systems, taking into account the angular resolution, sky coverage, and depth of LSST, and estimate the joint posterior of cosmological parameters over all events by MCMC sampling.

\subsection{Mock GW events}

We target GW events detected by the XG detector network that consists of an ET with the triangular configuration and two CEs. We adopt the orientation and the location of ET implemented in \texttt{Bilby}\footnote{https://git.ligo.org/lscsoft/bilby/-/blob/d37609ccc9a878750c841b4b93da990e63acffb2/bilby/gw/detector/\\detectors/ET.interferometer}, with a sensitivity curve for the 10 km arm length design\footnote{https://apps.et-gw.eu/tds/ql/?c=16492/ET10kmcolumns.txt}. We also assume that the two CE detectors locate at the LIGO Livingston site and the Hanford site with the same orientation as LIGO detectors, and a sensitivity curve for the 40 km arm length design\footnote{https://dcc.cosmicexplorer.org/CE-T2000017/public/cosmic\_explorer\_strain.txt}. We implement this in the Python tool \texttt{GWSim} \cite{Karathanasis:2022hrb} for mock GW generation. 

We first create unlensed GW catalogues for 5-year observation and 10-year observation respectively, assuming a duty cycle of 75\% for each detector (it applies to three ET detectors as a whole). We simulate both binary black hole (BBH) merger and neutron-star-black-hole (NSBH) merger events. For black hole mass distribution, we use the "Powerlaw + Double Peak" model adopted in the GWTC-4 cosmology paper \cite{LIGOScientific:2025jau}, while for neutron star we assume a uniform mass distribution model between 1.0 and 2.5 solar masses. We also adopt the Madau–Dickinson model as the merger rate redshift evolution model \cite{LIGOScientific:2025jau,Madau:2014bja}. We only simulate BBHs and NSBHs, because the event rate for binary neutron stars (BNSs) remains highly uncertain ($7.6-250~{\rm Gpc}^{-3}~{\rm yr}^{-1}$ from GWTC-4 \cite{LIGOScientific:2026vua}). A detailed description of these models and the injection values used is provided in Appendix \ref{app:model}. 

The luminosity distances of the GW events in the CMB rest frame are first derived from redshift assuming the fiducial $\Lambda$CDM model with $H_0 = 67.7\ {\rm km\ s^{-1}\ Mpc^{-1}}$ and $\Omega_{m,0} = 0.308$, based on the Planck 2018 TT, TE, EE+lowE+lensing+BAO results \cite{Planck:2018vyg}. Next, we add the cosmic dipole effect to the observed luminosity distances $d_{L}^{\rm obs}$ and observed masses of GW events based on the measurements from the CMB and the number count method respectively. In the first scenario, we adopt the observer's velocity as $v_o=370~{\rm km~s}^{-1}$ measured from the CMB, which corresponds to a cosmic dipole magnitude of $g=1.23\times10^{-3}$, with a direction of $(264^\circ,48^\circ)$ in the galactic coordinate. As for the second scenario, we assume the observer's velocity to be $v_o=800~{\rm km~s}^{-1}$, since the number count measurements yield roughly $600$-$1000~{\rm km~s}^{-1}$ in different studies. It corresponds to a cosmic dipole magnitude of $g=2.67\times10^{-3}$. We also assume the same cosmic dipole direction as the CMB measurement. In summary:
\begin{itemize}
    \item Scenario 1: CMB measurement. $g=1.23\times10^{-3}$, $(l^{\rm dip},b^{\rm dip})=(264^\circ,48^\circ)$.
    \item Scenario 2: Number count measurement. $g=2.67\times10^{-3}$, $(l^{\rm dip},b^{\rm dip})=(264^\circ,48^\circ)$.
\end{itemize}

\subsection{Lensed event selection}
\label{sec:lens_select}

We then simulate lensed events with the optical depth for mock GW sources following \cite{Sereno:2011ty,Chen:2026htz}. The optical depth describes how the probability of a GW event being lensed and detected varies with redshift. (see detailed derivation in Appendix \ref{app:optical_depth}). The optical depth $\tau$ is given by
\begin{equation}
    \tau = \frac{F_*}{30}[D_S(1+z_S)]^3 y_{\rm max}^2 \bigg\{ 1-\frac{1}{7}\frac{\Gamma[(8+\alpha)/\beta]}{\Gamma[(4+\alpha)/\beta]}\frac{\Delta t_*}{T_{\rm obs}} \bigg\},
    \label{eq:tau}
\end{equation}
where
\begin{equation}
    F_* = 16\pi^3 n_{*,0} \bigg(\frac{\sigma_{*,0}}{c} \bigg)^4 \frac{\Gamma[(4+\alpha)/\beta]}{\Gamma[\alpha/\beta]};
\end{equation}
\begin{equation}
    \Delta t_* = 32\pi^2 \bigg(\frac{\sigma_{*,0}}{c} \bigg)^4 \frac{D_S}{c}(1+z_S)y_{\rm max}.
\end{equation}
Here $\Gamma$ is the Euler gamma function, and $T_{\rm obs}$ is the observation time of GW detectors. The modified Schechter function parameters take the values of $n_{*,0}=8.0\times10^{-3}~h^3{\rm Mpc}^{-3}$, $\sigma_{*,0}=144~{\rm km}~{\rm s}^{-1}$, $\alpha=2.49$, and $\beta=2.29$, following previous study in \cite{Sereno:2011ty,Chen:2026htz}
% from the Sloan Digital Sky Survey (SDSS) Data Release 5 in \cite{Choi:2006qg} 
(also consistent with recent studies, e.g. in \cite{Choi:2006qg,Geng:2021tiz,2025MNRAS.537..779F}).
The maximal impact parameter $y_{\rm max}$ depends on the network SNR detection threshold $\rho_{\rm th}$ of the lensed GW images and the network SNR of the unlensed GW signal $\rho_{\rm UL}$:
\begin{equation}
    y_{\rm max} = \bigg[\bigg( \frac{\rho_{\rm th}}{\rho_{\rm UL}} \bigg)^2+1 \bigg]^{-1}.
    \label{eq:y_max}
\end{equation}
This choice of $y_{\rm max}$ ensures that all simulated lensed events satisfy $y\leqslant y_{\rm max}$, meaning that their second images are above the detection threshold. For events with $y\rightarrow 1$, the second image is suppressed and has a lower SNR. However, the sub-threshold search techniques enable the recovery of lensed signals with SNRs as low as 6, and occasionally down to 4 \cite{McIsaac:2019use, Li:2019osa, LIGOScientific:2023bwz, Li:2023zdl, LIGOScientific:2025cwb}, significantly enhancing the prospects for detecting strongly lensed signal pairs. Therefore, in our simulation, we compute $y_{\rm max}$ for each event using its unlensed network SNR $\rho_{\rm UL}$, and adopt a network SNR threshold of $\rho_{\rm th}=6$ for sub-threshold searches. Lensed events are then randomly selected from the mock GW catalogs with probability weighted by the optical depth in equation (\ref{eq:tau}) using the corresponding $y_{\rm max}$ for each event, and the exact value of $y$ for each lensed event is then drawn from a linear probability distribution $P(y)=y$ over the interval $(0, y_{\rm max})$.

For the selected lensed events, we generate foreground galaxy lenses following the stellar mass distribution of galaxies. The galaxy number density as a function of stellar mass is described by a double-Schechter function:
\begin{equation}
    \Phi(M) = \exp \bigg(-\frac{M}{M^*} \bigg) \bigg[\Phi^*_1 \bigg(\frac{M}{M^*}\bigg)^{\alpha_1+1} +\Phi^*_2 \bigg(\frac{M}{M^*}\bigg)^{\alpha_2+1} \bigg].
    \label{eq:double_schechter}
\end{equation}
We adopt the parameter values from \cite{2016MNRAS.459.2150W}: $\log (M^*/M_\odot)=10.79$, $\log (\Phi^*_1/h^3{\rm Mpc}^{-3})=-3.31$, $\log (\Phi^*_2/h^3{\rm Mpc}^{-3})=-2.01$, $\alpha_1=-1.69$, and $\alpha_2=-0.79$. Stellar masses for lens galaxies are randomly drawn with probabilities weighted by equation (\ref{eq:double_schechter}).
Given that the baryon-to-dark matter ratio in the universe is approximately 1:5, and that baryonic matter in galaxies includes both stars and interstellar medium, we adopt a representative stellar mass fraction of $10\%$ of the total galaxy mass (including the dark matter halo). The total lens galaxy mass $M_L$ is therefore taken to be ten times the drawn stellar mass. Lens redshifts $z_L$ are drawn with probabilities proportional to the differential comoving volume $dV_c/dz$.

\begin{figure}
    \centering
    \includegraphics[width=0.8\linewidth]{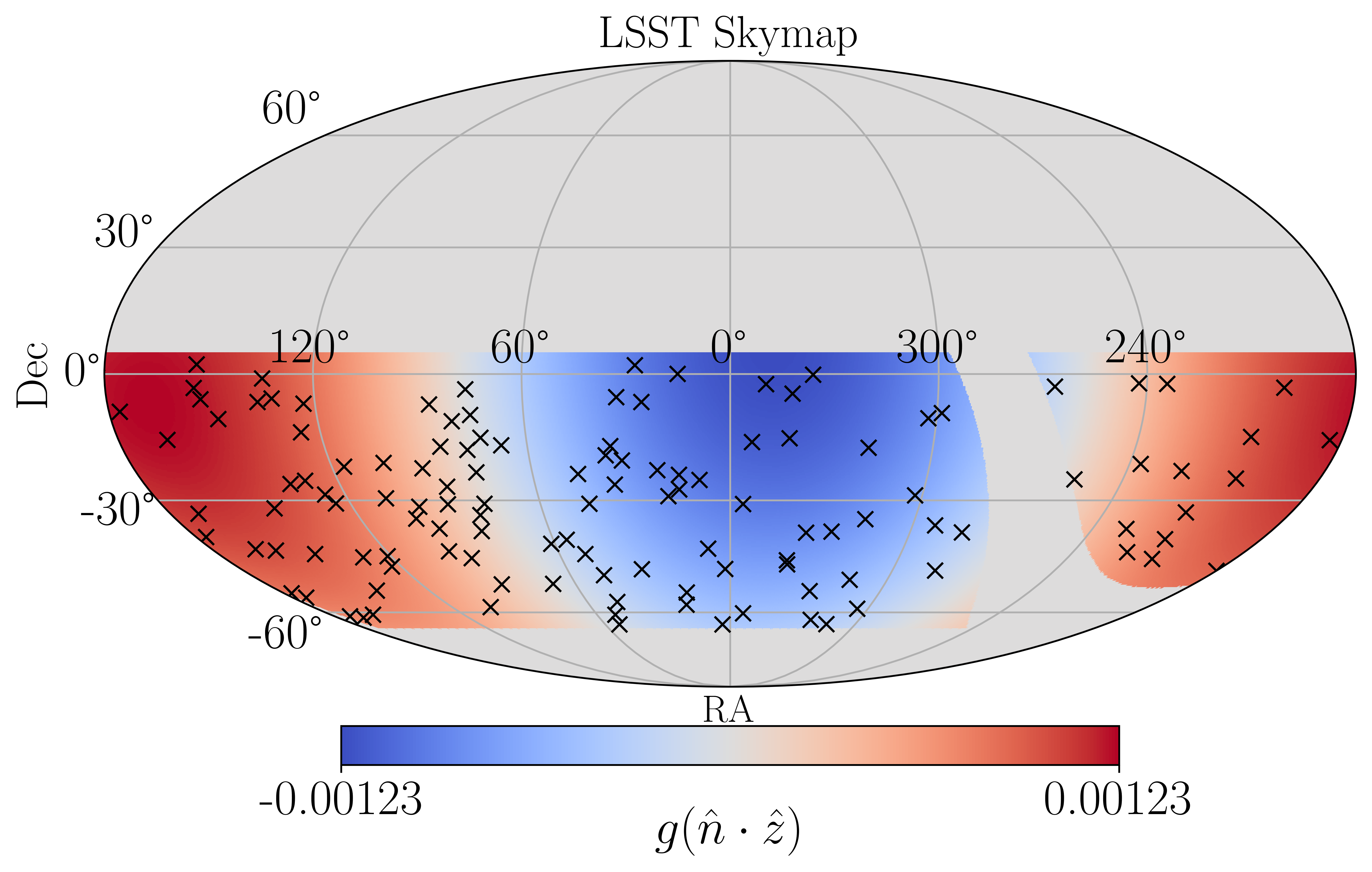}
    \caption{Skymap masked by LSST footprint for simulated lensed GW events (marked by $\times$) from 5-year observation of ET+CE whose host EM lensed images are resolvable by LSST, in a cosmic dipole background measured by the CMB.}
    \label{fig:skymap}
\end{figure}

Our method requires resolved images of lensed host galaxies, so we only select events whose sky locations are within galaxy survey's footprint and host galaxy images are resolvable by telescopes. Specifically, we consider the LSST survey, which will reach redshifts up to $z=3$ over 18,000 ${\rm deg}^2$ \cite{2009arXiv0912.0201L, LSSTDarkEnergyScience:2012kar}, covering most of the southern hemisphere of the sky. As shown in Fig. \ref{fig:skymap}, We assume that the LSST footprint will cover the area within $-65^\circ<{\rm RA}<5^\circ$, excluding the Milky Way bar in the center of the galactic coordinate with length of $160^\circ$ and width of $10^\circ$ \cite{2009arXiv0912.0201L}. In addition, the LSST camera has a resolvable angular resolution of 0.2 arc seconds \cite{LSST_overview}. For each selected lensed event, we compute $\theta_E$ from equation (\ref{eq:theta_E_sis}) using $M_L$ and the distances $D_L$, $D_S$, $D_{LS}$, which are obtained from $z_L$ and $z_S$ in the CMB rest frame. Events are considered resolvable if $\theta_\pm=(y\pm1)\theta_E>0.2''$. The time delay between images is then calculated using equations (\ref{eq:t_delay_dimensionless}), (\ref{eq:theta_E_sis}), and (\ref{eq:sis_time_delay}) with lens model parameters.

Applying the selection criteria described above, we identify 159 qualified lensed events within the LSST footprint for a simulated 5-year observation with the ET+CE network, including 95 BBHs and 64 NSBHs. Fig. \ref{fig:skymap} presents the sky map of these events, masked by LSST coverage and plotted against the CMB-measured cosmic dipole background. Meanwhile, extending the observation to 10 years yields 363 qualified events. Among all events, we randomly select $70\%$ to be doubly lensed events and $30\%$ to be triply/quadruply lensed events \cite{Li:2018prc}.

\subsection{Waveform parameter estimation}
\label{sec:waveform_PE}

For the selected doubly lensed events, we perform joint waveform parameter estimation (PE) for the two signal images in order to find the true $d_L^{\rm obs}$ and the impact parameter $y$, as shown in our previous work \cite{Chen:2026htz}. For the individual image, the magnification effect of lensing would lead to biases in the estimation of masses and luminosity distance. Using the SIS model, we can estimate $y$ from the magnification ratio of the two images of a lensed event, and thus find the true masses and $d_L^{\rm obs}$ in joint PE. For simplicity, we generate spin-free waveforms with the \texttt{IMRPhenomXPHM} waveform model \cite{Pratten:2020ceb}, and assume precise localization for the lensed events. So we perform joint PE for 5 parameters: observed chirp mass ${\cal M}_c^{\rm obs}$, mass ratio $q$, observed luminosity distance $d_L^{\rm obs}$, inclination angle $\iota$, and the impact parameter $y$. The higher modes in the \texttt{IMRPhenomXPHM} waveform would help break the degeneracy between $d_L^{\rm obs}$ and $\iota$. Furthermore, we assume that the Morse phase difference between the two image signals can be measured from the individual PE of the images prior to the joint PE, so we simply use a fixed Morse phase of $\pi/2$ between the images.

In our joint PE, the posterior of waveform parameters $\{\phi\}$ given image data $x_{j=1,2}$ is given by
\begin{equation}
    P(\phi|x_1,x_2) = \int P(x_1,x_2|\phi) \Pi(\phi) {\rm d}\phi,
\end{equation}
where $\Pi(\phi)$ is the prior, and the joint log likelihood is given by
\begin{equation}
    \log P(x_1,x_2|\phi) = -\frac{1}{2} \sum_{j=1}^2 \langle \tilde{h}(x_j)-\tilde{h}(\phi)| \tilde{h}(x_j)-\tilde{h}(\phi) \rangle.
\end{equation}
Here $f$ is the frequency, $\tilde{h}(f)$ is the detector response to GW signals, given by 
\begin{equation}
    \tilde{h}(f)= \sum_k [F_+^k \tilde{h}_+(f) + F_\times^k \tilde{h}_\times(f)],
\end{equation}
where $F_+^k$ and $F_\times^k$ are the antenna pattern functions of the $k$ detector for the plus and cross GW polarization. The scalar product $\langle \tilde{h}(f)|\tilde{g}(f) \rangle$ is defined as 
\begin{equation}
    \langle \tilde{h}(f)|\tilde{g}(f) \rangle \equiv 2~{\rm Re}\bigg[\int {\rm d}f \frac{\tilde{h}(f)\tilde{g}^*(f)+\tilde{h}^*(f)\tilde{g}(f)}{{\rm Sn}(f)} \bigg],
\end{equation}
where ${\rm Sn}(f)$ is the detector noise power spectral density (PSD). We adopt priors $\Pi(d_L^{\rm obs}) \propto (d_L^{\rm obs})^2$ and $\Pi(y)\propto y$, while uniform priors are used for all other parameters. Finally, posterior samples are obtained using \texttt{nessai} \cite{Williams:2021qyt,Williams:2023ppp}, which implements normalizing flows within nested sampling as part of the \texttt{Bilby} package \cite{bilby_paper}.

\begin{figure}[t!]
    \centering
    \includegraphics[width=0.9\linewidth]{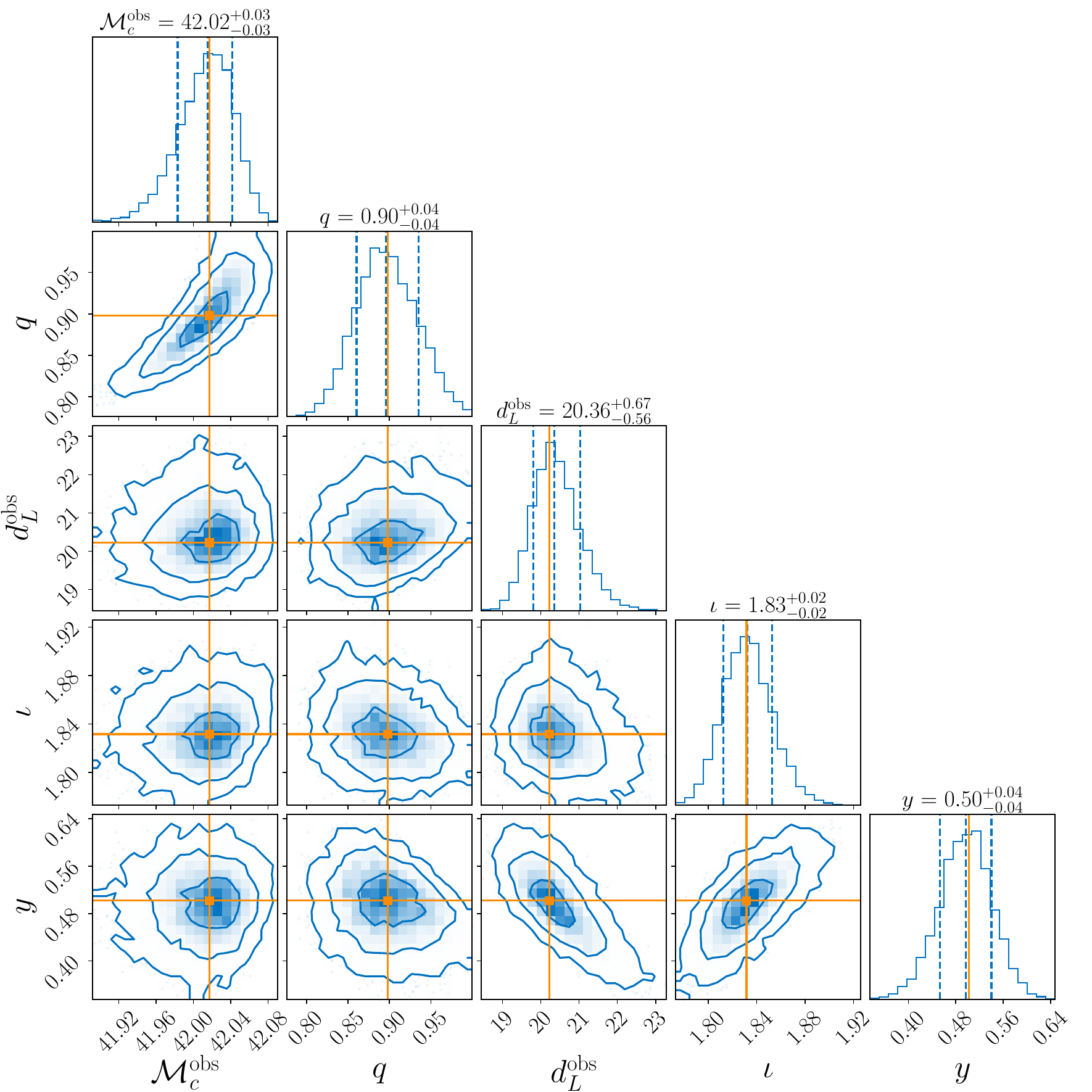}
    \caption{Posterior samples from joint re-analysis of one of the mock doubly lensed BBHs detected by ET+CE network using \texttt{nessai} nested sampling in \texttt{Bilby}. The estimated parameters include chirp mass ${\cal M}_c[M_\odot]$, mass ratio $q$, GW luminosity distance $d_L^{\rm obs}$[Gpc], inclination angle $\iota$, and impact parameter $y$. The orange lines show the injected values for parameters.}
    \label{fig:waveform_PE}
\end{figure}

Fig. \ref{fig:waveform_PE} presents an example of the posterior samples obtained from joint PE for a pair of lensed GW signals with SNR of 52.1 and 30.0 respectively. Despite moderate degeneracies, the true values of $d_L^{\rm obs}$ and $y$ are well recovered, and their marginalized posteriors approximately follow Gaussian distributions. Accordingly, we can extract the observables and their uncertainties from the mean and standard deviation of the posterior samples, respectively.

\section{Results}
\label{sec:results}

We use the \texttt{emcee} package \cite{Foreman-Mackey:2012any} to infer the posterior distributions of cosmological parameters by maximizing the log likelihood given in equation (\ref{eq:likelihood}), adopting uniform priors on all parameters. We perform the inference in both scenarios where the cosmic dipole is injected based on CMB measurements and the number count measurements respectively. For each scenario, we forecast the constraints expected from the ET+CE network after 5 years and 10 years of observation. The results are presented below.

\subsection{Double systems}

\begin{figure}[t!]
    \centering
    \includegraphics[width=0.7\linewidth]{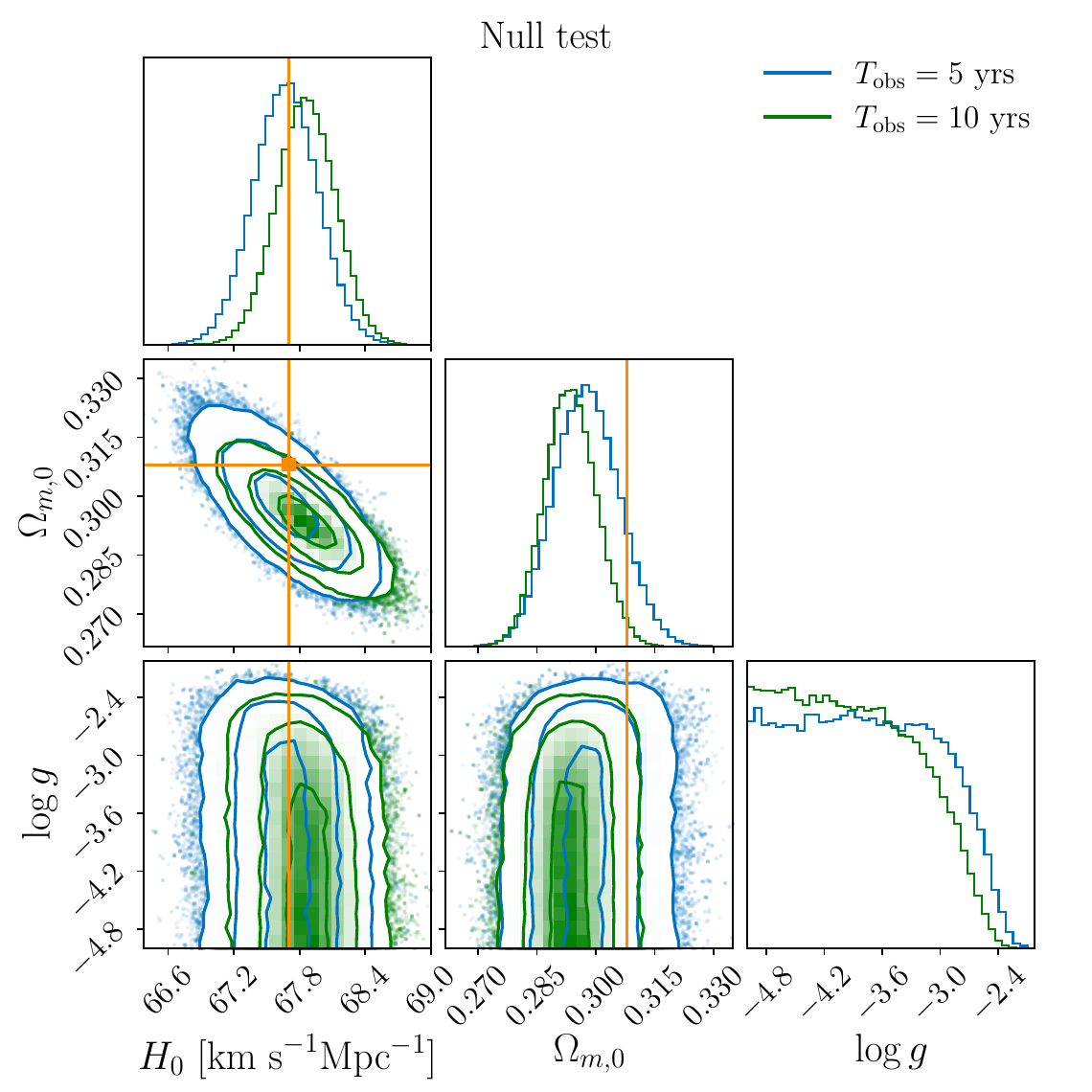}
    \caption{Joint likelihood forecast for $H_0$, $\Omega_{m,0}$ and $\log g$ in the null test where $g=0$, assuming 5-year and 10-year observation by ET+CE respectively. The orange lines show the injected parameter values.}
    \label{fig:corner_null}
\end{figure}

We first forecast joint posteriors on $H_0$, $\Omega_{m,0}$, and the dipole magnitude $g$, with fixed dipole direction using simulated doubly lensed events in both cosmic dipole scenarios, as well as a null test where $g=0$. The prior on $g$ is uniform within $(10^{-5},1)$.
The inferred joint likelihood for the null test with 5-year and 10-year observation are shown in Fig. \ref{fig:corner_null}. For clearer visualization, we plot the likelihood of $\log g$ instead of $g$. As shown in the corner plot, the marginalized likelihood of $\log g$ has a clear upper bound, and the likelihood extends to lower value of $\log g$. In addition, the upper bound for 10-year observation is lower than that for 5-year observation, indicating a stronger constraint on $\log g$. The results are in good agreement with the injection in which no cosmic dipole is present.

\begin{figure}[t!]
    \centering
    \includegraphics[width=0.7\linewidth]{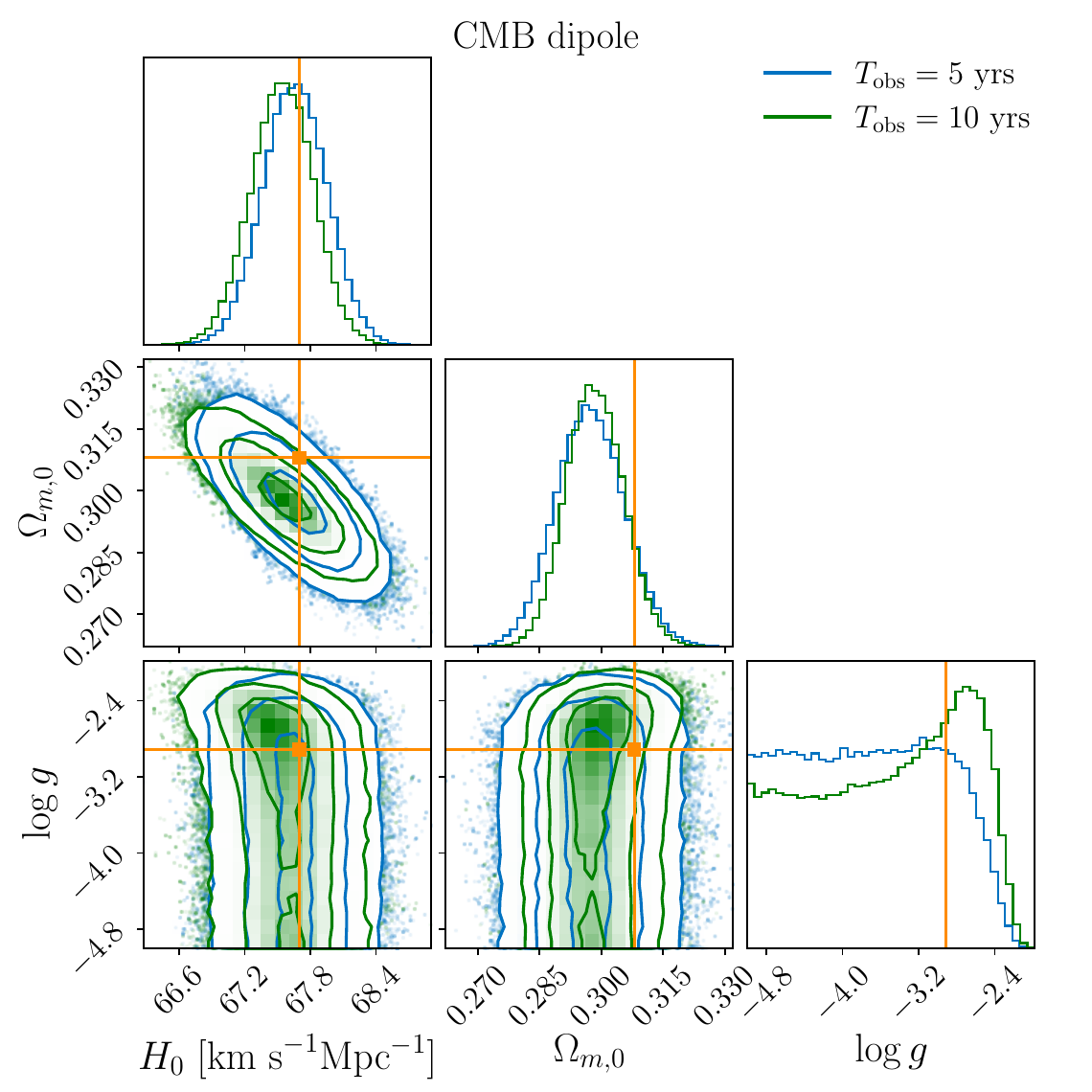}
    \caption{Joint likelihood forecast for $H_0$, $\Omega_{m,0}$ and $\log g$ with fixed dipole direction in the CMB dipole scenario, assuming 5-year and 10-year observation by ET+CE respectively. The orange lines show the injected parameter values.}
    \label{fig:corner_CMB}
\end{figure}

Next, we present the inferred joint likelihood for 5-year and 10-year observation in the CMB dipole scenario in Fig. \ref{fig:corner_CMB}. We can see that the upper bound of $\log g$ is also well constrained, while the lower bound of $\log g$ is still unconstrained. Considering the $90\%$ confidence interval with the lower bound fixed at the prior lower bound, the upper bound of the marginalized likelihood of $\log g$ corresponds to $g=1.94\times10^{-3}$ for 5 years of observation, which is $\sim58\%$ larger than the true vale. For 10 years of observation, there exists a peak in the marginalized likelihood distribution of $\log g$, where the peak and the $1\sigma$ confidence interval around it gives $\log g=-2.73^{+0.75}_{-1.18}$. The peak, the upper bound and the lower bound of the posterior of $g$ correspond to $g=1.86\times10^{-3}$, $g=1.04\times10^{-2}$, and $g=1.24\times10^{-4}$, but the lower bound is mainly driven by the prior range.

In addition, $H_0$ and $\Omega_{m,0}$ are tightly constrained, with an uncertainty of $\sim0.43\%$ and $\sim2.8\%$ respectively for 5 years of observation, which are consistent with our forecasts for the $\Lambda$CDM model in \cite{Chen:2026htz}. The mean value slightly shifts away from the injected value, likely due to degeneracy, but is within the 2$\sigma$ contour. In particular, they have little correlation with $g$. For 10 years of observation, the uncertainty of $H_0$ and $\Omega_{m,0}$ slightly decrease to $\sim0.41\%$ and $\sim2.3\%$ respectively.

\begin{figure}[t!]
    \centering
    \includegraphics[width=0.7\linewidth]{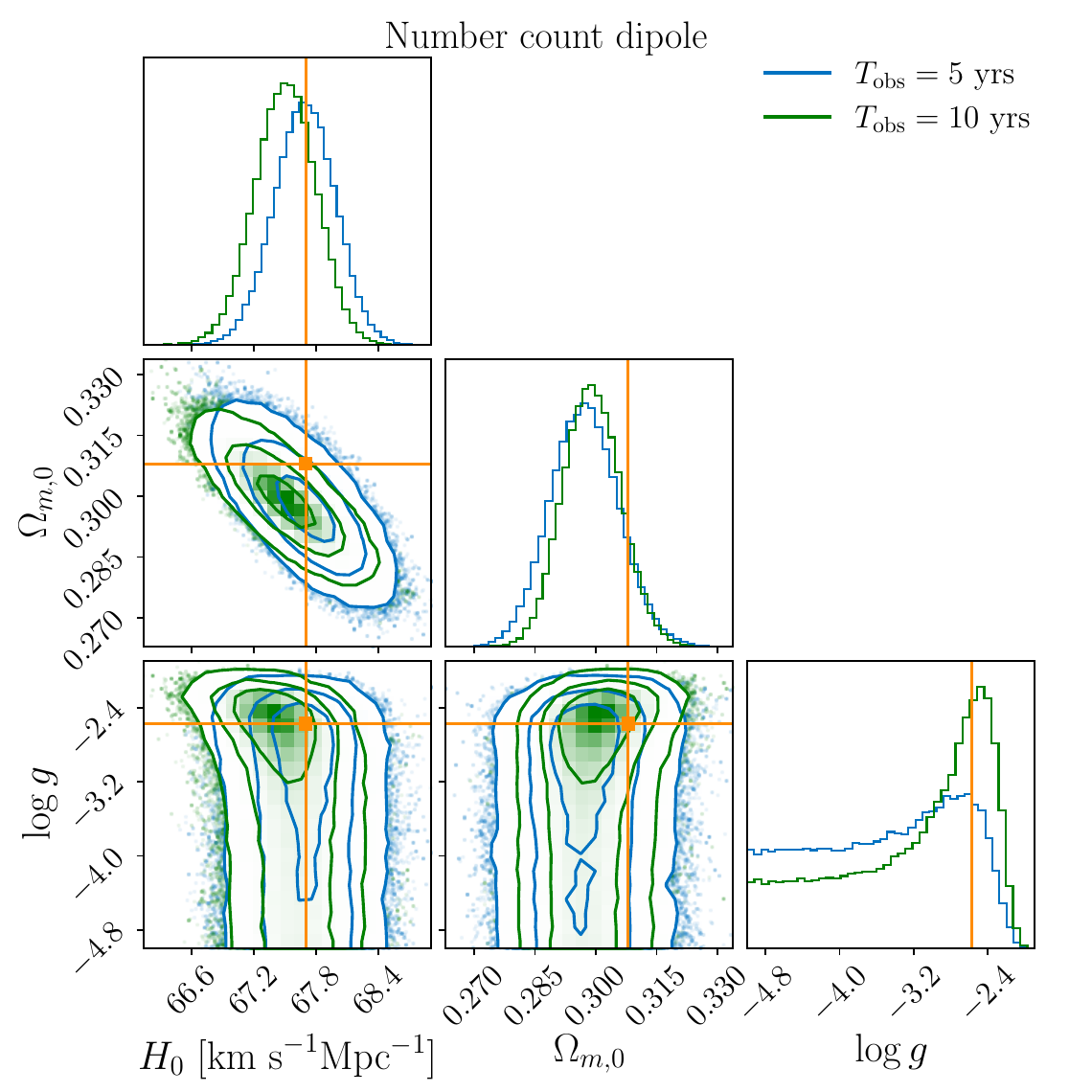}
    \caption{Joint likelihood forecast for $H_0$, $\Omega_{m,0}$ and $\log g$ with fixed dipole direction in the number count dipole scenario, assuming 5-year and 10-year observation by ET+CE respectively. The orange lines show the injected parameter values.}
    \label{fig:corner_NC}
\end{figure}

On the other hand, in the number count dipole scenario, the constraint on $g$ is generally stronger than the CMB dipole scenario, because the number count dipole magnitude is about twice larger. As shown in Fig. \ref{fig:corner_NC}, the marginalized likelihood of $\log g$ results in a peak around the injected value for both 5-year and 10-year observation periods. Whereas, similar to the CMB dipole scenario, the likelihood is bounded only at the upper limit, while the lower limit remains unbounded. Quantitively, the peak and the $1\sigma$ confidence interval centered at the peak yield $\log g=-2.75^{+0.78}_{-1.15}$ for 5-year observation, and $\log g=-2.55^{+0.66}_{-0.90}$ for 10-year observation, respectively. While the constraint for 5-year observation is less accurate, for 10-year observation, the peak of the posterior of $g$ given a flat prior corresponds to $g=2.79\times10^{-3}$, with the upper and lower bound to be $g=1.27\times10^{-2}$ and $g=3.53\times10^{-4}$. 

These results show that doubly lensed GWs + galaxy catalogue with the SIS model could potentially measure the cosmic dipole given a longer period of observation, and the true value is recovered better for the number count dipole scenario. But the uncertainty on $g$ leads to an upper bound nearly an order of magnitude larger than the true value, and it couldn't rule out the scenario for $g\rightarrow0$. On one hand, the error in $d_L^{\rm obs}$ estimated from GWs for distant lensed events could be up to $\sim 10\%$, making it difficult to probe the tiny modification in $d_L^{\rm obs}$ from the cosmic dipole. On the other hand, the time-delay constraint relies on accurate measurement on $y$, which is also estimated from joint GW analysis with uncertainty of $\sim5-10\%$. Thus, the cosmic dipole effect is hard to measured from time delay constraint with the SIS model as well.

\subsection{Combining triples and quadruples}

Apart from doubly lensed events, there are also 30\% of the total lensed events with more than two images, mainly triples and quadruples, which can provide more accurate lens reconstruction, and hence tighter cosmological constraint by time delay measurement with equation (\ref{eq:chi_square_quad}), replacing the second part in equation (\ref{eq:chi_square}). Performing the same inference with \texttt{emcee} for triple/quadruple events randomly selected from our mock lensed event pool, we obtain joint likelihood of $H_0$, $\Omega_{m,0}$, and $\log g$ as for the double events. Then we combine the joint likelihood for double events and triple/quadruple events by multiplying the multidimensional likelihood from the two analyses, and obtain the marginalized likelihood of $\log g$ from the combined likelihood.

\begin{figure}[t!]
    \centering
    \begin{subfigure}{0.6\linewidth}
    \includegraphics[width=\linewidth]{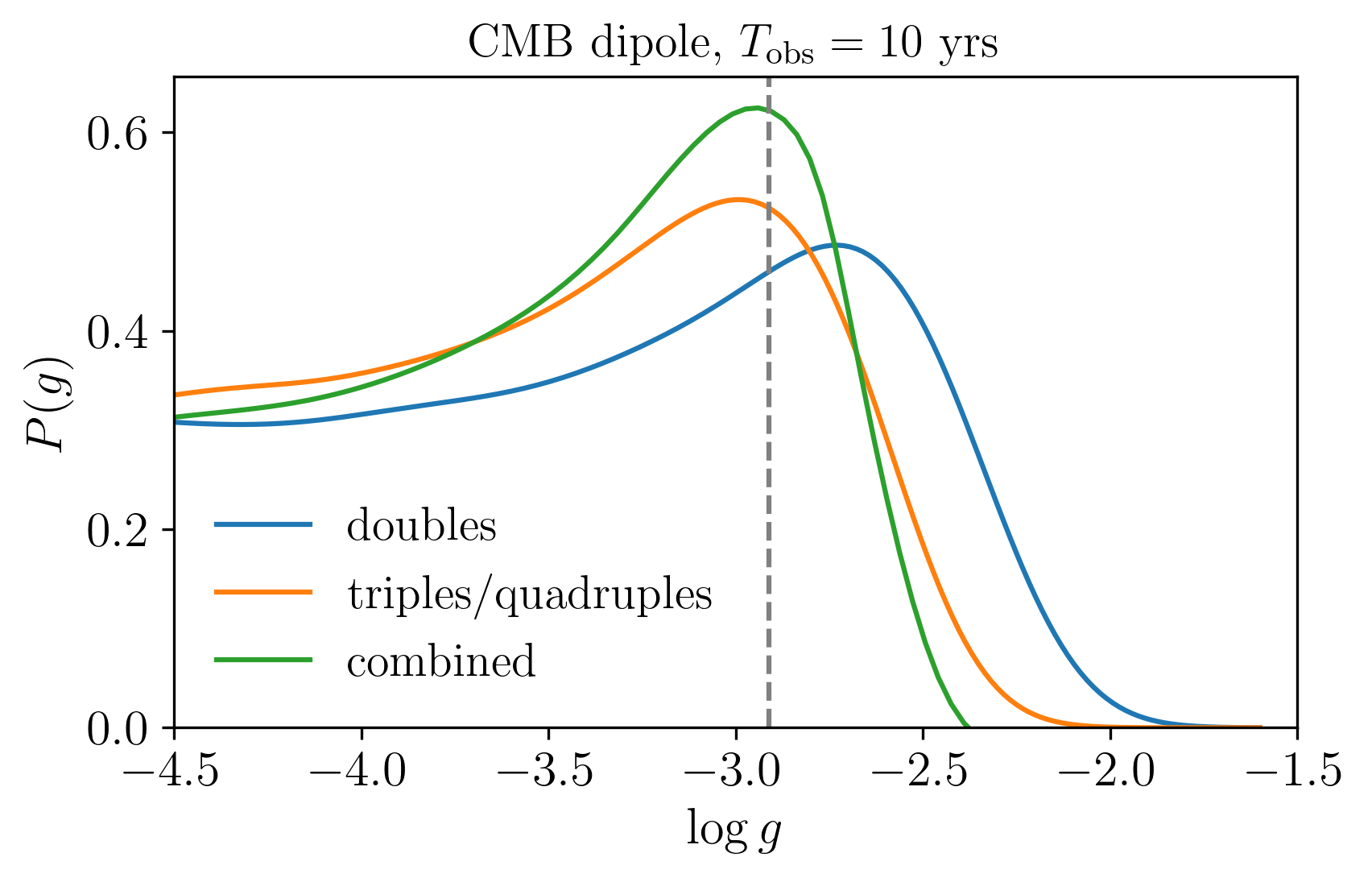}
    \end{subfigure}
    \begin{subfigure}{0.6\linewidth}
    \includegraphics[width=\linewidth]{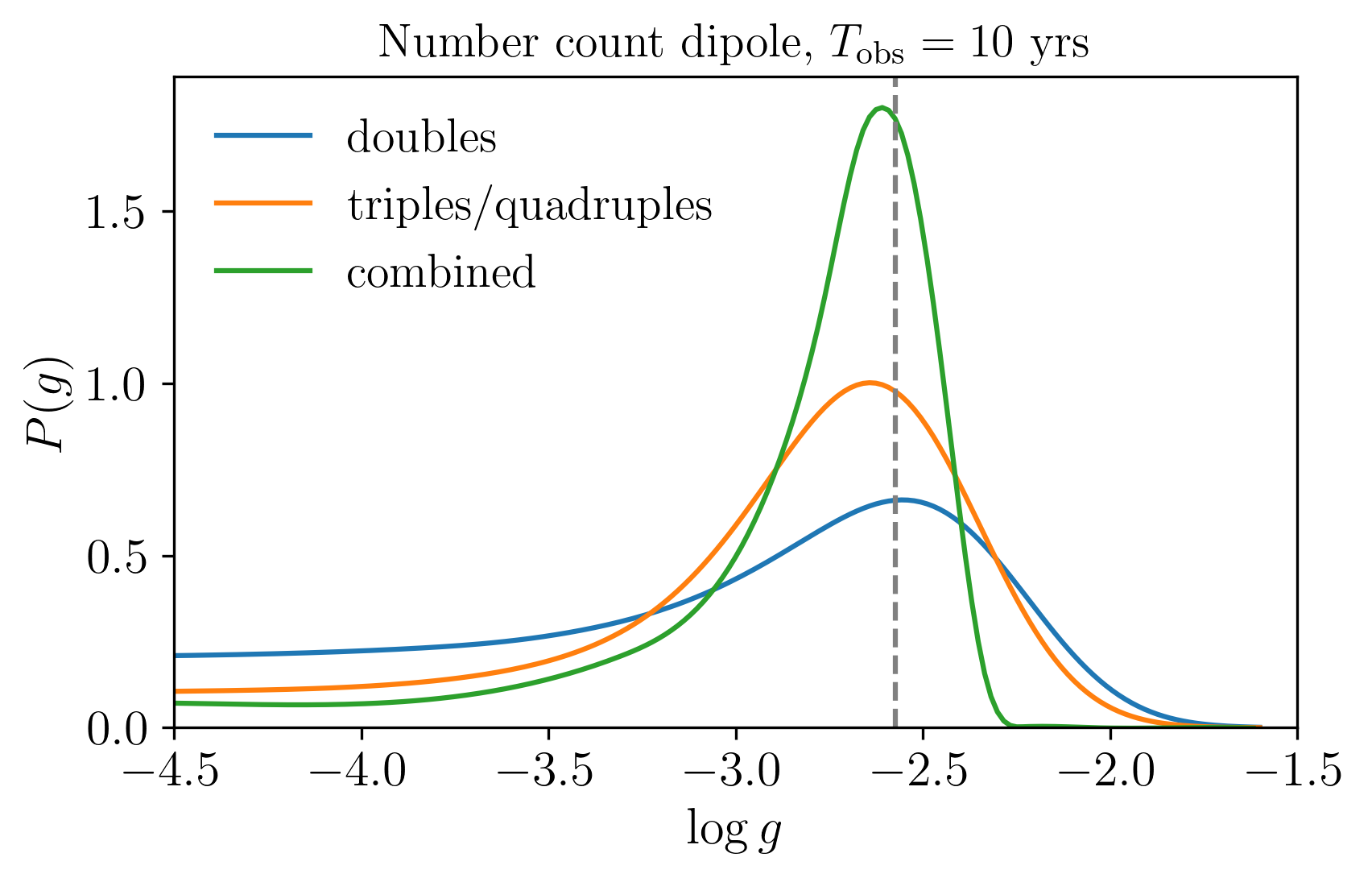}
    \end{subfigure}
    \caption{Posterior distribution of $g$ from individual likelihood for doubly lensed events (blue) and triply/quadruply lensed events (orange) respectively, and from combined likelihood (green), for 10-year observation in the CMB dipole (upper panel) and the number count dipole (lower panel) scenario. The injected values of $g$ are shown in gray dashed lines. }
    \label{fig:combined_constraint_quad}
\end{figure}

Fig. \ref{fig:combined_constraint_quad} shows the posterior of $g$ from analysis for double events and triple/quadruple events respectively, as well as the combined posterior for them, in both the CMB dipole scenario and the number count dipole scenario for 10 years of observation. Firstly, we find that the constraint on $g$ by triple/quadruple events is slightly stronger than that by double events in both scenarios, despite the fact that the number of triple/quadruple events are more than a half lower. Such improvement is driven by smaller uncertainty in lens reconstruction for time-delay cosmography. Secondly, the combined constraint on $g$ for double events and triple/quadruple events is significantly improved. For the CMB dipole scenario, the peak of the combined posterior and the $1\sigma$ confidence interval around it gives $\log g=-2.95^{+0.28}_{-1.18}$. The peak corresponds to $g=1.12\times10^{-3}$, which is closer to the injected value. The $1\sigma$ upper bound corresponds to $g=2.14\times10^{-3}$, which is about $74\%$ larger than the injected value. However, the lower bound is still not able to be constrained. Meanwhile, for the number count dipole scenario, the peak of the combined posterior and the $1\sigma$ confidence interval gives $\log g=-2.61^{+0.21}_{-0.32}$. The peak corresponds to $g=2.45\times10^{-3}$, and the $1\sigma$ upper bound corresponds to $g=3.98\times10^{-3}$, which is $62\%$ larger than the injected value. More importantly, although the posterior is still unbounded at lower $g$, the tail of the posterior at lower $g$ becomes much smaller. The $1\sigma$ lower bound yields $g=1.17\times10^{-3}$, which is $56\%$ smaller than the injected value.

\begin{figure}[t!]
    \centering
    \includegraphics[width=0.7\linewidth]{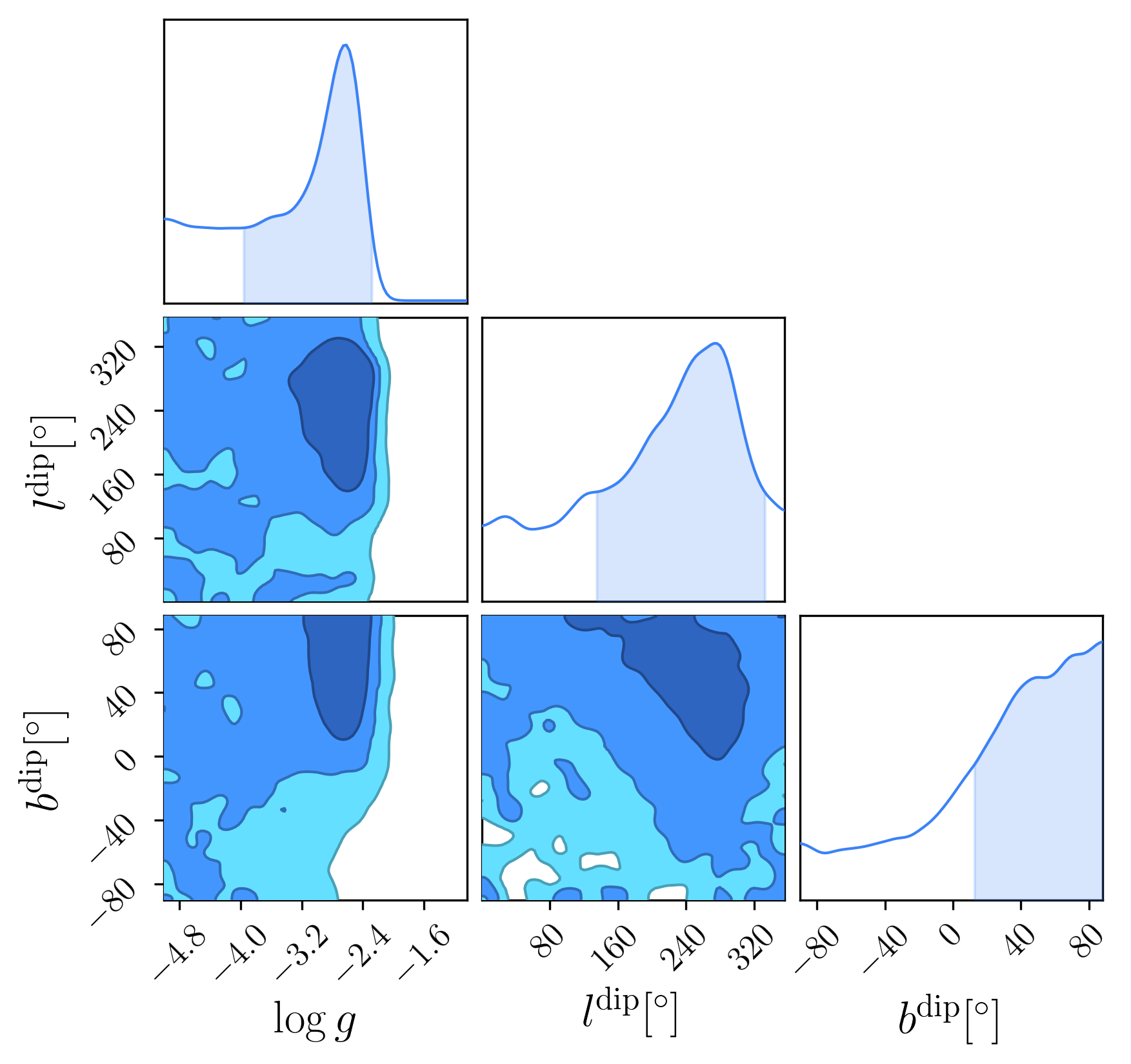}
    \caption{Joint likelihood of dipole magnitude $g$ and dipole direction $(l^{\rm dip}, b^{\rm dip})$ by combining likelihood for double and triple/quadruple events in the number count dipole scenario with 10 years of observation.}
    \label{fig:corner_NC_dipole}
\end{figure}

In addition to the dipole magnitude, we also investigate the constraint on the dipole direction in the galactic coordinate, $l^{\rm dip}$ and $b^{\rm dip}$, with the combined likelihood of double and triple/quadruple events for 10 years of observation. The constraints on $H_0$ and $\Omega_{m,0}$ are only weakly affected when $l^{\rm dip}$ and $b^{\rm dip}$ are included in the inference, whereas $g$ is strongly affected by it. For the CMB dipole scenario, the peak in the posterior of $g$ from the combined likelihood disappears, and the distribution is generally flat starting from the prior lower bound and drop sharply at $\log g\lesssim-3$. Moreover, $l^{\rm dip}$ and $b^{\rm dip}$ are very weakly constrained. On the other hand, constraints on the dipole are significantly stronger for the number count dipole scenario, as shown in Fig. \ref{fig:corner_NC_dipole}. The peak and the $1\sigma$ confidence interval of the marginalized likelihood of $g$ yield $\log g=-2.57^{+0.24}_{-1.17}$, which has a larger uncertainty compared to the case when fixing the dipole direction. Meanwhile, there exists a moderate constraint on $l^{\rm dip}$, with the peak and the $1\sigma$ bound to be $l^{\rm dip}=(264.2^{+66.7}_{-107.6})^\circ$. Whereas, the posterior of $b^{\rm dip}$ peaks at the prior upper bound of $b^{\rm dip}=90^\circ$, with the $1\sigma$ lower bound to be $b^{\rm dip}=16.9^\circ$. Therefore, even in the most optimistic case of 10-year observation on the number count dipole, the joint constraint on cosmic dipole magnitude and direction is weaker than the number count measurement with traditional EM probes.

\subsection{Discussion}

In summary, our forecast shows that associating strongly lensed GWs with the LSST galaxy catalogue can potentially measure the cosmic dipole magnitude given a long observation period of ET+CE like 10 years, but it is more difficult to constrain the dipole direction at the same time. In particular, the constraint of the dipole is stronger if the true dipole magnitude is larger, as in our number count dipole scenario compared to the CMB dipole scenario. In the most optimistic case of 10-year observation in the number count dipole scenario, our forecasted constraint gives $g=(2.45^{+1.53}_{-1.28})\times10^{-3}$ while fixing the dipole direction, corresponding to $\sim57\%$ uncertainty. Such uncertainty is larger than the forecast with number count of $10^5$ GW events by ET and CE within 1 year of observation \cite{Grimm:2023tfl}. Given the same 10-year observation period by ET and CE, the number count method can measure $g$ with $\sim10\%$ uncertainty. The same precision could also be achieved by 1000 BNS bright sirens in the era of next-generation detectors \cite{Cai:2019cfw,Cousins:2024bhk}. Moreover, using the golden dark siren detected by ET+CE, $g$ can be constrained with $6\%$ uncertainty in joint measurement with the dipole direction. While the constraint from combining strongly lensed GWs and a galaxy catalogue is not the tightest, it nevertheless offers an independent measurement of the cosmic dipole as a cross-check, and could provide a unique insight into the cosmic dipole tension.

On the other hand, the limitation of our method also lies in the redshift range of galaxy catalogues. ET and CE are expected to detect GW events up to $z=10-20$. Since strong lensing can magnify GW signals, events with redshift beyond the detector horizon could also potentially be observed if the lensed images are magnified. More importantly, the lensing probability of events is significantly increased at higher redshift. Therefore, future large-scale galaxy survey is the key to increase the number of galaxy-catalogue associated strongly lensed GWs, and thus to enhance our constraint on the cosmic dipole. For example, assuming a galaxy catalogue with redshift range up to $z=5$ and a sky coverage of $90\%$ to be achieved by multiple observatories like the MegaMapper, we repeat our simulation and obtain 1496 lensed GW events that can be associated with galaxy catalogues for 10 years of observation. Considering $\Delta g$ to be scaled by $1/\sqrt{N}$, based on the constraint with 363 events associated with LSST, we expect to reduce the uncertainty to $\sim 28\%$. In addition, we could obtain stronger constraint on the dipole direction as well. 

Compared to other GW measurement methods, the core innovation of our method is to probe the cosmic dipole along with other cosmological constants by combining the dipole modification to GW luminosity distance, galaxy redshift, and the time-delay distance in strong lensing for the first time. Given the impact of the cosmic dipole to strong-lensing cosmography with quasars \cite{Dalang:2023usx}, it is also important to avoid biases by the cosmic dipole effect in time delay cosmography with strongly lensed GWs. Our forecast explicitly shows that $H_0$ and $\Omega_{m,0}$, and potentially more parameters in alternative cosmology models, can be measured precisely in joint estimation with the cosmic dipole without strong correlation. However, since we use the SIS model to approximate the doubly lensed system, systematics inherent to the lens model would degrade the precision of cosmological constraints, as discussed in \cite{Chen:2026htz}, so a more comprehensive lens model will be required to obtain more robust constraints. Furthermore, the distortion of the EM lensed images by the angular aberration effect would also contribute to probing the cosmic dipole through Einstein radius modification \cite{Millon:2026nwo}. Integrating such effect in our method in future works could potentially strengthen the constraint of the cosmic dipole.

\section{Conclusions}
\label{sec:conclusions}

The primary challenge for GW cosmology is measuring redshifts of GW events. Strong lensing not only enables time-delay cosmography, but also greatly enhances GW localization, allowing host galaxy redshifts to be determined by cross-matching with lensed galaxy catalogues.
Our work demonstrates the potential of measuring the cosmic dipole by associating strongly lensed GW events with galaxy catalogues. 
% In particular, we combine the constraints from doubly lensed systems with the SIS model and from triply or quadruply lensed systems assuming a precise lens potential reconstruction. 
We forecast the constraint on the dipole magnitude $g$ jointly with $H_0$ and $\Omega_{m,0}$ in the $\Lambda$CDM model using mock strongly lensed GW events including BBHs and NSBHs that are within the LSST footprint and resolvable by the LSST camera, but no real or mock galaxy catalogue is used. 
Given the cosmic dipole tension, we forecast the constraint in the CMB dipole scenario and the number count dipole scenario respectively. 
% We fix the dipole direction at $(264^\circ,48^\circ)$ in the galactic coordinate, and infer the joint likelihood of cosmological parameters by MCMC sampling assuming 5 years or 10 years of observation by a detecting network consisting ET and two CEs. 

We find that using doubly lensed events with the SIS model alone, the CMB dipole magnitude can only be probed with 10 years of observation, while the number count dipole magnitude can be measured for both 5-year and 10-year observation when fixing dipole direction, but with large uncertainty. Then we enhance the constraints on $g$ for 10 years of observation by combining the joint likelihood from doubly lensed events and that from triply/quadruply lensed events assuming a precise lens potential reconstruction. For the CMB dipole scenario, the posterior of $g$ peaks at $g=1.12\times10^{-3}$, with $1\sigma$ upper uncertainty of $74\%$ but no lower bound. For the number count dipole scenario, the tail of the posterior $P(g)$ extending toward lower values is substantially reduced, and we obtain $g=(2.45^{+1.53}_{-1.28})\times10^{-3}$. However, when jointly measure the dipole direction, the constraint on $g$ is weakened, and the dipole direction is poorly constrained. 

While not reaching the accuracy of cosmic dipole measurements obtained with other EM probes, our method provides a complementary and independent means of measurement. The dipole constrained by our method will be a tracer of late-time Universe properties, which is expected to be consistent with the number count measurements. In this case, our method would support the possibility of new physics operating between the early and late universe that leads to the cosmic dipole tension. But if our method yields a more consistent result with the CMB, the tension would then be more likely to arise from unknown systematics inherent to the number count approach. Therefore, our method could offer a potential path toward resolving the cosmic dipole tension.

On the other hand, several challenges remain in our method for achieving a precise measurement of the cosmic dipole. First, this work presents merely an idealized forecast, serving as a proof of principle for the measurement of the cosmic dipole via the association of strongly lensed GWs with galaxy catalogues. We imply observation conditions of LSST on mock strongly lensed GWs without generating a mock galaxy catalogue, and assume that we identify the true EM lens images associated with GWs. But in reality, one may need to marginalize likelihood of lens parameters over different lens systems within the GW localization area, and compare Bayes factors for different lenses. In addition, a more concrete Bayesian inference also needs to consider selection effects in terms of event SNR, galaxy survey magnitude threshold, magnification bias, etc. A more comprehensive study with an actual mock galaxy catalogue needs to be done to include all the mentioned effects in our next step. 
Second, the precision of time-delay cosmography depends on the accuracy of the lens model we adopt for lens reconstruction. For doubly lensed events, the lens galaxy ellipticity and the mass-sheet degeneracy would result in biases in time-delay cosmography with the SIS model. While a more careful choice would be to adopt a lens model more complex than the SIS model for doubly lensed events, such as the singular isothermal ellipsoid model, this would increase degeneracies between lens model parameters and cosmology, ultimately leading to weaker cosmological constraints. Moreover, the lens images in EM observations can display irregular morphologies (e.g., arcs), which complicates the extraction of a well-defined image position angle, and thus degrades the cosmological constraints. Last but not least, we have neglected the peculiar velocity of the lens and the source galaxies in our simulation, which would lead to systematic biases in our cosmic dipole measurement. Future study is necessary to eliminate the effects introduced by peculiar velocities in a realistic galaxy catalogue.

Nevertheless, future studies aimed at improving the robustness of lens reconstruction from GW signal images are essential for obtaining accurate cosmological constraints with our method. Furthermore, the cosmic dipole constraint derived from our approach can be strengthened by next-generation galaxy surveys that offer a larger redshift range, broader sky coverage, and enhanced resolution. By providing an independent cross-check of the number count method, our approach will contribute to resolving the cosmic dipole tension---one of modern cosmology's most intriguing anomalies---over the coming decades.

\begin{acknowledgments}
A.C. is supported by the China Postdoctoral Science Foundation under Grant No. 2025M773325, and the National Natural Science Foundation of China (NSFC) under Grant No. E414660101 and 12147103. J.Z. is supported by the NSFC under Grants No.~E414660101 and No.~12147103, and the Fundamental Research Funds for the Central Universities under Grants No.~E4EQ6604X2 and No.~E3ER6601A2. We are grateful to the High Performance Computing Center (HPCC) of ICTP-AP for performing the numerical computations in this paper.
\end{acknowledgments}

\appendix

\section{GW population model}
\label{app:model}

In our simulation, BBH events are generated with a population model in mass and redshift space given by
\begin{equation}
    p_{\rm pop}(m_1^s,m_2^s,z|\Lambda_m,\Lambda_r) \propto p(m_1^s,m_2^s|\Lambda_m)p_{\rm rate}(z|\Lambda_r)\frac{{\rm d}V_c}{{\rm d}z},
\end{equation}
where $m_1^s,m_2^s$ are the source-frame mass. For the black hole mass distribution model, we use the Power-law + Double Peak phenomenological model that provides a good fit to the GWTC-4 data \cite{LIGOScientific:2025jau}. This model is characterized by 10 parameters: the minimum and maximum mass $M_{\rm min}$ and $M_{\rm max}$; the power-law slope $\alpha_p$ of the primary mass distribution; the fraction $\lambda_{\rm g}$ of the two Gaussian components; the fraction $\lambda_{\rm g,low}$ of the lower Gaussian component; the means $\mu_{\rm g,low}$ and $\mu_{\rm g,high}$ of the lower and higher Gaussian component; their widths $\sigma_{\rm g,low}$ and $\sigma_{\rm g,high}$; and the tapering range $\delta_m$ at the lower end of the mass distribution. The primary mass prior for this population model is thus given by
\begin{align}
    p(m_1^s|\Lambda_m) & = (1-\lambda_g){\cal B}(m_1^s|M_{\rm min},M_{\rm max},\alpha_p) \nonumber\\
    & +\lambda_g\lambda_{\rm g,low}{\cal G}(m_1^s|\mu_{\rm g,low},\sigma_{\rm g,low}) \nonumber\\
    & +\lambda_g(1-\lambda_{\rm g,low}){\cal G}(m_1^s|\mu_{\rm g,high},\sigma_{\rm g,high}),
\end{align}
where ${\cal B}$ denotes the broken power-law distribution, and ${\cal G}$ the Gaussian distribution. The secondary mass is then drawn from a power-law mass ratio distribution with slope $\beta_p$.

\begin{figure}
\centering
\begin{subfigure}{0.48\textwidth}
\includegraphics[width=\textwidth]{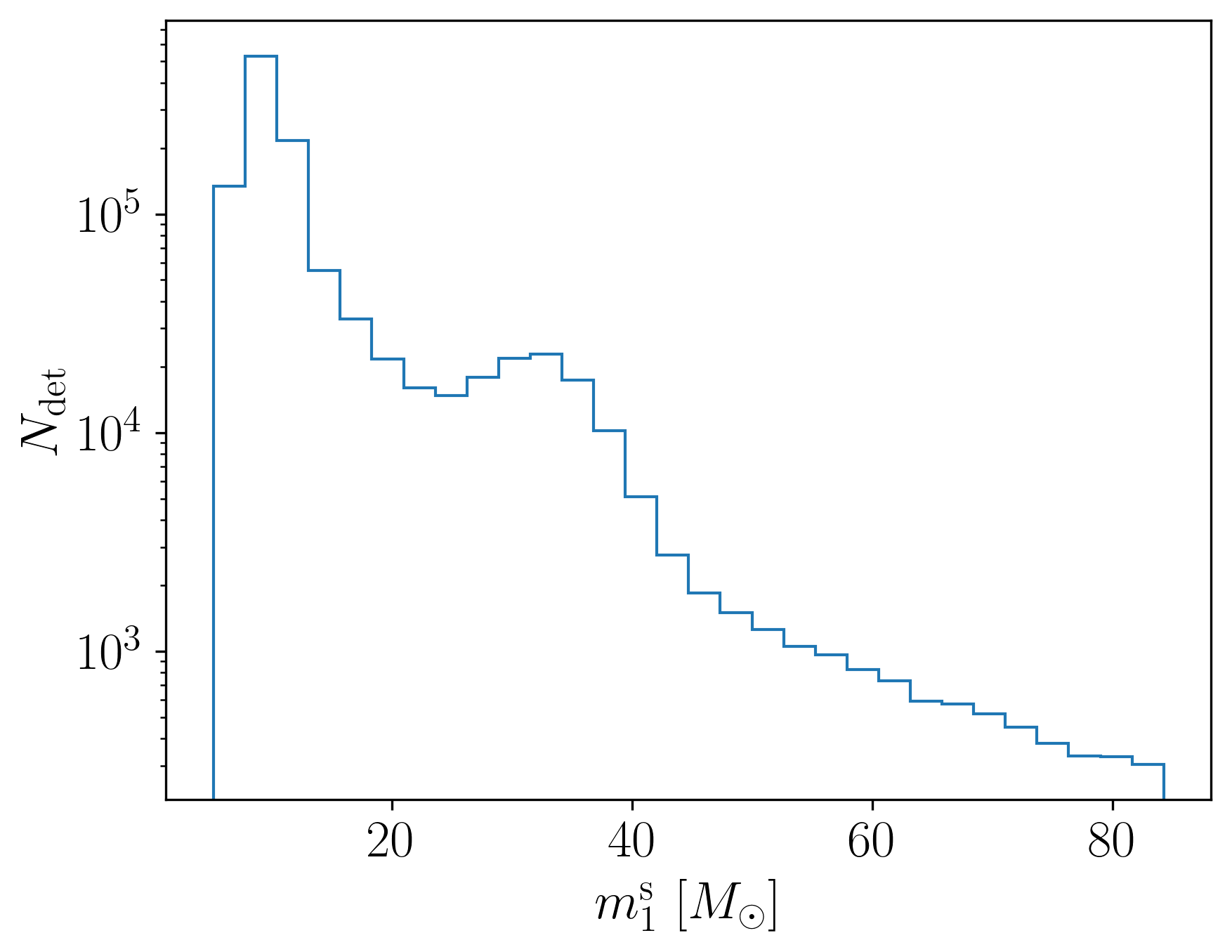}
\end{subfigure}
\begin{subfigure}{0.497\textwidth}
\includegraphics[width=\textwidth]{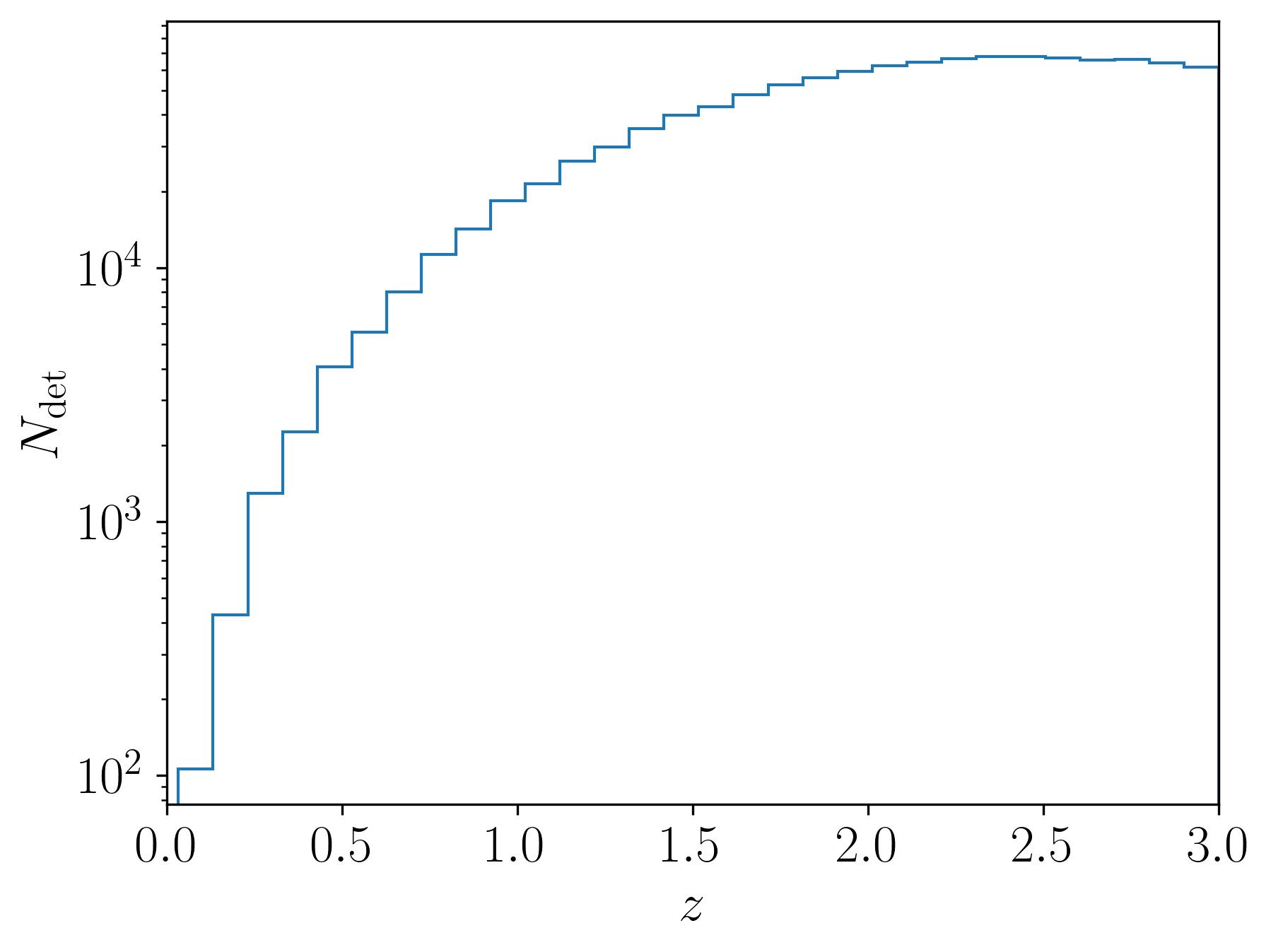}
\end{subfigure}
\caption{Mock detected event number of ET+CE for 5 years with SNR threshold of 6. Upper panel: primary source mass distribution with the power-law + double-peak model. Lower panel: redshift distribution combing uniform-in-comoving-volume density and Madau-Dickinson merger rate evolution model. }
\label{fig:Pm1_Pz}
\end{figure}

We also incorporate the Madau–Dickinson merger rate redshift evolution model, inspired by the Madau–Dickinson star formation rate \cite{Madau:2014bja}, into our simulation. This model is also used in the GWTC-4 cosmology study \cite{LIGOScientific:2025jau}. The merger rate as a function of redshift is given by
\begin{equation}
    p_{\rm rate}(z|\Lambda_r) = R_0 [1+(1+z_p)^{-\gamma-k}] \frac{(1+z)^\gamma}{1+\left(\frac{1+z}{1+z_p}\right)^{\gamma+k}},
\end{equation}
where $R_0$ is the merger rate at $z=0$, and $\gamma$ and $k$ are the power-law slopes of the rate evolution at redshift below and above the turning point $z_p$, respectively. In our simulation, we adopt the parameter values from the GWTC-4 cosmology analysis \cite{LIGOScientific:2025jau}, as listed in Table \ref{tab:GWTC4_param}. We also adopt $R_0=20~{\rm Gpc}^{-3}{\rm yr}^{-1}$ for BBHs and NSBHs, which is consistent with GWTC-4 population study \cite{LIGOScientific:2026vua}. We show the corresponding primary source mass distribution and redshift distribution for our mock detected events by ET+CE with an SNR threshold of 6 in Fig. \ref{fig:Pm1_Pz}.
\renewcommand{\arraystretch}{1.5}
\begin{table}
    \centering
    \caption{Parameters for the Power-law + Double Peak black hole mass distribution model and the Madau-Dickinson merger rate redshift evolution model used in our mock event simulation.}
    \label{tab:GWTC4_param}
    \begin{tabular}{c|c|c|c|c|c|c}
        \hline
        $M_{\rm min}[M_\odot]$ & $M_{\rm max}[M_\odot]$ & $\alpha_p$ & $\beta_p$ & $\mu_{\rm g,low}[M_\odot]$ & $\sigma_{\rm g,low}[M_\odot]$ & $\lambda_{\rm g,low}$  \\
        \hline
        $4.72$ & $84.28$ & $~2.97~$ & $~0.83~$ & $9.91$ & $0.83$ & $0.86$ \\
        \hline
        \hline
        $\mu_{\rm g,high}[M_\odot]$ & $\sigma_{\rm g,high}[M_\odot]$ & $\lambda_{\rm g,high}$ & $\delta_m[M_\odot]$ & $\gamma$ & $k$ & $z_p$ \\
        \hline
        $32.31$ & $4.24$ & $0.28$ & $4.71$ & $~3.46~$ & $~2.90~$ & $~2.77$ \\
        \hline
    \end{tabular}
\end{table}

\section{Optical depth of sources}
\label{app:optical_depth}

% \begin{figure}
% \centering
% \includegraphics[width=0.49\textwidth]{fig_ML_pML.png}
% \caption{Stellar mass density profile of lens galaxies derived from the double Schechter function in equation (\ref{eq:double_schechter}).}
% \label{fig:phi_M}
% \end{figure}

The probability that a background source is gravitationally lensed by a foreground galaxy is referred to as the optical depth. Its differential form with respect to the lens redshift and velocity dispersion is given by \cite{Sereno:2011ty}
\begin{equation}
    \frac{{\rm d}^2\tau}{{\rm d}z_L{\rm d}\sigma} = \frac{{\rm d}n}{{\rm d}\sigma}(z_L, \sigma)s_{\rm cr}(z_L, \sigma) \frac{c{\rm d}t}{{\rm d}z_L}(z_L),
    \label{eq:differential_tau}
\end{equation}
where $s_{\rm cr}$ denotes the cross section for signal deflection in the lens system. The number density distribution of lenses as a function of velocity dispersion $\sigma$ can be modeled by a modified Schechter function \cite{SDSS:2003fyb}:
\begin{equation}
    \frac{{\rm d}n}{{\rm d}\sigma} = n_{*}\bigg(\frac{\sigma}{\sigma_*} \bigg)^{\alpha} \exp\bigg[-\bigg(\frac{\sigma}{\sigma_*} \bigg)^{\beta} \bigg] \frac{\beta}{\Gamma[\alpha/\beta]}\frac{1}{\sigma},
\end{equation}
where $\alpha$ is the faint-end slope, $\beta$ controls the high-velocity cut-off, and $n_{*}$ and $\sigma_*$ are the characteristic number density and velocity dispersion. These can be parameterized to evolve with redshift as $n_*(z)=n_{*,0}(1+z)^{3-\nu_{n^*}}$ and $\sigma_*(z)=\sigma_{*,0}(1+z)^{\nu_{\sigma^*}}$. In this work, we assume a constant comoving lens number density, i.e., $\nu_{n^*}=\nu_{\sigma^*}=0$.

\begin{figure}[t]
\centering
\includegraphics[width=0.6\textwidth]{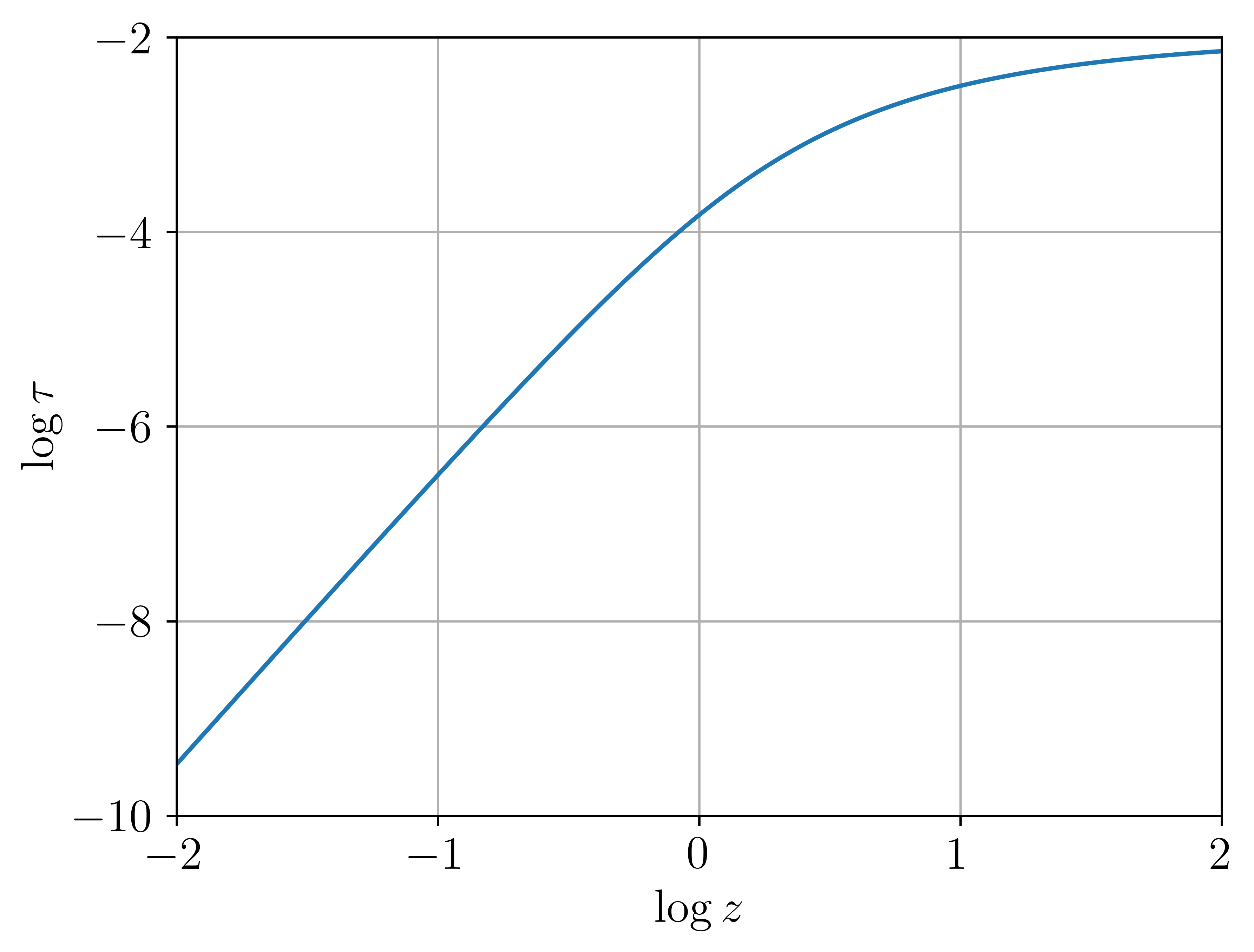}
\caption{Optical depth of background sources computed using equation (\ref{eq:tau_z}) with parameters $n_{*,0}=8.0\times10^{-3}~h^3{\rm Mpc}^{-3}$, $\sigma_{*,0}=144~{\rm km}~{\rm s}^{-1}$, $\alpha=2.49$, and $\beta=2.29$.}
\label{fig:tau_z}
\end{figure}

For a finite-duration transient observation, the lensing statistic must account for missing events due to time delays between lensed signals \cite{Oguri:2002hv}. The cross-section for a singular isothermal sphere (SIS) model in a transient survey is
\begin{equation}
    s_{\rm cr}(z_L, \sigma) = \pi D_L^2\theta_E^2 \int_{y_{\rm min}}^{y_{\rm max}} f(\Delta t(y;\sigma,z_L))y{\rm d}y,
    \label{eq:cross_section}
\end{equation}
where $f(\Delta t)$ is the observable fraction of lensed events with time delay $\Delta t$. Assuming a uniform arrival time distribution of lensed signals over the continuous observation period $T_{\rm obs}$, we have
\begin{equation}
    f(\Delta t) = 1-\frac{\Delta t}{T_{\rm obs}}
\end{equation}
for $\Delta t<T_{\rm obs}$, and $f(\Delta t)=0$ otherwise. Equation (\ref{eq:cross_section}) then yields
\begin{equation}
    s_{\rm cr} = \pi D_L^2\theta_E^2 \bigg[(y_{\rm max}^2-y_{\rm min}^2)-\frac{2}{3}\frac{\Delta t_z}{T_{\rm obs}}(y_{\rm max}^3-y_{\rm min}^3) \bigg].
\end{equation}
For the SIS model, $y_{\rm min}=0$. Substituting the above expression into equation (\ref{eq:differential_tau}) and integrating over $\sigma$ from $0$ to $\infty$, we obtain
\begin{equation}
    \tau = \frac{F_*}{30}[D_S(1+z_S)]^3 y_{\rm max}^2 \bigg\{ 1-\frac{1}{7}\frac{\Gamma[(8+\alpha)/\beta]}{\Gamma[(4+\alpha)/\beta]}\frac{\Delta t_*}{T_{\rm obs}} \bigg\},
    \label{eq:tau_z}
\end{equation}
where
\begin{equation}
    F_* = 16\pi^3 n_{*,0} \bigg(\frac{\sigma_{*,0}}{c} \bigg)^4 \frac{\Gamma[(4+\alpha)/\beta]}{\Gamma[\alpha/\beta]};
\end{equation}
\begin{equation}
    \Delta t_* = 32\pi^2 \bigg(\frac{\sigma_{*,0}}{c} \bigg)^4 \frac{D_S}{c}(1+z_S)y_{\rm max}.
\end{equation}
Figure \ref{fig:tau_z} displays the optical depth computed using the parameters from Section \ref{sec:lens_select}. 
% Additionally, Fig. \ref{fig:phi_M} shows the stellar mass density profile of lens galaxies, modeled by the double Schechter function in equation (\ref{eq:double_schechter}).

\bibliography{reference}{}
\bibliographystyle{JHEP}

\end{document}